\documentclass[journal]{IEEEtran}

\usepackage{amsmath}
\usepackage{enumitem}
\usepackage{hhline}
\usepackage{graphicx}
\usepackage{multirow}
\usepackage{color}
\usepackage{fancybox}
\usepackage{cite}
\usepackage{url}
\usepackage[normalem]{ulem}
\usepackage{siunitx}
\usepackage{algorithm}
\usepackage{subfig}
\usepackage{breqn}
\usepackage{booktabs}

\usepackage{xcolor,etoolbox}
\usepackage{soul}
\usepackage{clipboard}

\definecolor{gray1}{gray}{0.7}
\definecolor{gray2}{gray}{0.98}
\definecolor{light-gray}{gray}{0.95}

\newcommand{\ignore}[1]{}
\newcommand{\red}[1]{\textcolor{red}{#1}}

\newcommand{\gray}[1]{\textcolor{gray1}{#1}}

\begin{document}
\bstctlcite{IEEEexample:BSTcontrol}

\title{Toward Multi-kW Power Delivery Methodologies for Advanced 3D Heterogeneous Integration}

\author{Peiyi Yue, Hangyu Zhang, Ratul Das, Ramesh Harjani, and Sachin S. Sapatnekar
\thanks{The authors are with the University of Minnesota, Minneapolis, MN.  This effort is supported in part by the NSF under awards 2212345, 2324945, 2403408, and 2437795, and by the SRC JUMP2.0 CHIMES Center.}}

\maketitle

\begin{abstract}
The demands of modern applications require the construction of ever more complex integrated systems, with AI applications in particular serving as a significant driver for increased system size, potentially going beyond the trillion-transistor mark.  The path to building these systems requires the use of advanced packaging, with heterogeneous integrated 2D and 3D chiplets placed atop a substrate.  Such computationally powerful systems require significant power for computation: reliable and robust power delivery is a major challenge in light of high power densities and pin count bottlenecks.  This paper overviews approaches to building design methodologies that overcome this problem, through the design of multistage distributed power delivery systems, optimized for performance and reliability, and built to coexist within stringent performance, thermal, and reliability constraints.
\end{abstract}	


\ignore{
\noindent
\red{To-do:
\begin{itemize}
\item \gray{Abstract}
\item \gray{Conclusion}
\item \gray{Add more figures}
\item Insert Ratul's input
\item Check work by \gray{Madhavan Swaminathan}, \gray{Inna Partin-Vaisband}, \gray{Subu Iyer}, \gray{Kaladhar Radhakrishnan}
\item \gray{Check text overlap with SKL paper}
\item \gray{Check popular literature - Mark Swinnen, Ed Sperling}
\item \gray{Look at reports from ASIC Consortium discussions}
\item \gray{Clean up reference section}
\end{itemize}
}
}

\section{Introduction}
\label{sec:intro}

\noindent
The push towards greater integration is driven by the increasing demands of high-performance computation, artificial intelligence, and big-data applications that require large amounts of data movement.  For cloud applications, conventional monolithic computing systems with limited die sizes cannot service the performance needs of modern AI applications: for instance, it takes 20 NVIDIA H100 GPUs to hold one copy of the GPT4 model parameters~\cite{Dally23}, and data movement costs are very high. The slowdown in Moore’s law leaves a gap in satisfying the exponentially-growing hardware needs of advanced chips. For edge applications, bandwidth limitations from sensor/memory to compute modules pose fundamental challenges in the big data era: a 2-bit DVS camera with 4M pixels, uses a 1TBps data rate at 1 MHz~\cite{Krishnan23}; graph neural nets on the Nell dataset with 64K graph nodes require up to 2.7 TB of data to be communicated between processing elements~\cite{Mandal22}; large language models demand over TBps-level data rates~\cite{Wang25b}.

These drivers have led to tremendous interest in heterogeneous integration (HI)~\cite{HIR_2024}, a form of 3D integration that places multiple chiplets, each of which may be built using a planar 2D or 3D technology, on a substrate and connects them with dense, fine-pitch interconnects on organic, glass, or silicon interposer substrates.  HI provides a viable approach for increasing integration levels at low cost and high yield.  New HI interconnect technologies~\cite{Intel_EMIB, CoWoS_3D, Intel_Foveros} with through-silicon vias (TSVs)/through-glass vias (TGVs), microbumps, hybrid bonding, Cu-Cu bonding, and Cu pillars, have provided pathways for meeting the demands of modern and future data- and compute-intensive workloads. HI brings tremendous compute capability within a package, while also incorporating low-latency access to memory and limiting the costs of data movement~\cite{Lakefield_3D, Gomes22, Zeppelin_3D, AMD_3D, Agarwal22, INTACT_Vivet_Pascal, Pal21}.

Moreover, HI raises new opportunities for integrating 2D or 3D chiplets on a 2.5D substrate built in different technology nodes, or using disparate technologies, allowing each function to be implemented in a technology that best matches its performance, cost, and reliability requirements. For instance, cutting-edge digital systems typically use advanced technology nodes for lower power, while analog/mixed-signal (AMS) and radio-frequency (RF) components often remain in mature technology nodes (e.g., 65nm or 180nm) for improved voltage tolerance, better device matching, and lower cost. RF front ends may also leverage SiGe BiCMOS or GaAs-based processes to provide strong performance for specific use cases such as RF/mm-wave circuits and power amplifiers. In addition, HI facilitates the integration of high bandwidth memories (HBMs) and silicon photonic interconnects, overcoming the limitations of traditional electrical links. A representative HI system may integrate sensing (image sensors, phased arrays, and RF antennas), analog front-end processing, a digital baseband that provides the horsepower for on-chip sensing, low-latency signal processing, as well as on-chip and off-chip communication.

However, the task of performing a large amount of computing within a small footprint also brings together new challenges.  One of the chief issues is the problem of ensuring that sufficient power is supplied to the integrated system to support this additional computation.  It is projected that the power required by a 3D HI system will be considerably larger than that for today's most advanced chips, but will have to be supplied by a comparable number of pins.  Projections show that such systems will require multiple kiloWatts of power within the small footprint; as a point of comparison, a domestic refrigerator today consumes about half a kiloWatt, over a considerably larger footprint.

To deliver power to multi-kW HI systems with low losses, a strong focus must be placed on developing robust power delivery solutions that assure power integrity.  The severity of the problem goes well beyond the already-difficult problem of power delivery for 2D chips: predictions show that HI systems can experience sustained power densities of 1--10 W/mm$^2$~\cite{HIR_2024}, as opposed to power densities in the range of 0.1–0.5 W/mm$^2$ for traditional 2D systems. Beyond ensuring correct chip operation, this problem is critical from a global standpoint: as chip power consumption increases, it takes up a growing percentage of global energy consumption.  In 2024, datacenter power was reported to be at 4.4\% of US energy consumption, a number that is projected to grow to up to 12\% by 2028~\cite{DoE2023}. Achieving efficiencies that approach or exceed 90\% for these high-power integrated systems is therefore an imperative piece of the solution to controlling global energy consumption.

In this paper, we consider the problem of power delivery to a 3D HI system, considering the design of multiple stages of voltage regulators as well as the power distribution network (PDN) that delivers power to individual chiplets.  Specifically, we address issues related to the design of these networks and new design automation tools and methodologies required to ensure robust power delivery in a 3D HI system.  The magnitude of this problem is unprecedented in integrated system design: today's state-of-the-art die require about 700~W of power at about 1~W/mm$^2$; in contrast, future HI systems are projected to require multiple kW of power~\cite{Tang25} at densities running up to 1--10~W/mm$^2$.  The path to systematically solving these problems is still being mapped out, and this is a nascent field: there has been relatively little work on systematically examining the design space and enumerating the possible solutions.  As a first step, in~\cite{Krishnakumar26}, approaches for delivering 1--50kW of power to systems, using direct 48V-to-1V, 48V-to-24V-to-1V, and 48V-to-12V-to-1V, have been explored. Today's cutting-edge high-power systems may use a couple of GPUs in a package, where the scope of the power delivery problem can be limited due to the uniformity of the structure.  Future systems will be much more heterogeneous, as they are disaggregated into a diverse set of chiplets that share the same substrate~\cite{Nalla25}. To the best of our knowledge, no general solutions to this general problem have been proposed for the general problem of disaggregating large systems, which constitute the input to the power delivery problem. The aim of this paper is to draw from techniques that have been used on related design problems, such as on-chip power delivery, to outline the design space for 3D HI power delivery, and therefore to lay the ground for future work in this area. A recent complementary survey on this topic~\cite{Swaminathan26} addresses design and technology solutions to this problem; this paper focuses on modeling, analysis, and optimization solutions for design automation for vertical power delivery in HI systems.

The decisions that drive power delivery design must be made in conjunction with other parts of the HI design process. Building a large HI system requires (1)~methods for chiplet disaggregation that map the system to smaller 2D or 3D chiplets, working in conjunction with system-technology co-optimization (STCO) to determine reasonable design decisions that optimize computation and communication, together with the choice of substrate and chiplet technologies; (2)~procedures for multiphysics~\cite{Wang25} and multiscale analyses that incorporate thermomechanical aspects into performance analysis, ranging from fast machine-learning-driven analyses in early stages to signoff-quality multiphysics-based analysis; (3)~physical design techniques for placing and routing chiplets and embedded active/passive elements on and within the substrate, including the design of thermal and power delivery solutions; and (4)~the underlying infrastructure required to facilitate 3D HI design, including the design of chiplet libraries and the establishment of data formats and standards.   All of these are intertwined with decisions associated with power delivery design.

It is important to point out that a conventional approach could model the chiplet as the point of load (PoL) for power delivery, but chiplets may, or may not, have their own internal regulation mechanisms.  In fact, the actual PoL is at the switching elements, but this would require the solution of very large systems, neglecting the natural partition between the HI substrate and the chiplet.  We will discuss ways of bridging this gap by considering chiplet regulation capabilities while building the power delivery solution in the HI substrate. Some of the chief considerations that arise in the design of power delivery solutions for 3D HI systems are as follows:

\noindent 
\textbf{Generality of the methodology to multiple 3D HI technologies.}
Power delivery methodologies must be general enough to address the common problems of all HI systems, and yet specific enough to be applicable within the constraints of specific technologies. For instance, the HI substrate may support the introduction of only interconnects, or interconnects and passive devices, or interconnects and passive as well as active devices.  In the first case, the substrate supports merely the PDN and any regulators must be embedded within chiplets; in the second, it also supports decoupling capacitors and inductors that could be used by a regulator chiplet; and in the third, distributed regulators can be placed throughout the substrate.  This has a significant impact on the achievable efficiency and effectiveness: the cardinal rule of power delivery is that the closer the voltage regulator can be to the point of load, the more effective it can be.

\noindent
\textbf{Design of the high-level architecture for power delivery.}
The power delivery solution requires the development of a high-level architecture for power delivery, followed by detailed design.  A typical architecture includes a first-stage regulator that takes a high-voltage off-chip voltage supply and converts it to an intermediate bus voltage (IBV) for distribution to second-stage voltage regulator modules.  Depending on the HI technology, these modules may be distributed within the HI package to achieve optimal efficiency and to maintain effective voltage regulation in the presence of PDN losses.  

\noindent
\textbf{Integration with other 3D HI system optimizations.}
The design of the 3D HI chip goes through multiple stages, from early-stage to late-stage design. In early design, the system is disaggregated into multiple chiplets, e.g., by selecting chiplets from a library~\cite{Nalla25}.  These disaggregated chiplets must go through a placement stage.  In various early- and late-stage design steps, due to the high power densities, a power delivery solution must be codesigned at an appropriate level of granularity to serve the needs of the system. This requires fast analysis and optimization techniques for early-stage system design, integrated into the disaggregation or placement optimization, and more deliberate optimization as more detailed design information is available in late-stage system design. For example, during disaggregation, it is necessary to determine whether to choose a chiplet with on-chip regulators; or one with front-side power delivery, or both front-side and back-side connections; or to use a set of separate regulator chiplets.

\noindent
\textbf{Support for self-consistent iterative refinement from early-stage to late-stage design.}
The power delivery solution typically undergoes multiple rounds of refinement as the design is developed.  During early-stage design, key design decisions about the architecture of the solution must be made, depending on the degrees of freedom provided by the HI substrate and the limited information available at that stage.  These high-level decisions constrain and guide the subsequent implementation of the entire power delivery solution.  For example, early decisions on the power delivery architecture must be made using coarse, and sometimes no, knowledge of chiplet disaggregation or placement; these must be refined in a self-consistent manner throughout the design flow, ensuring that the detailed physical implementation remains compatible with earlier decisions.

\noindent
\textbf{Interactions with multiphysics and reliability issues.}
Components of the power delivery solution are embedded within the 3D HI stack, and like chiplets and the interconnects, face thermal and mechanical stresses that may alter performance~\cite{Avula22}. Elevated temperatures also exacerbate reliability risks and failures in devices and interconnects, which must be factored into the design of regulators and PDN interconnects.

The paper is organized as follows.  Section~\ref{sec:architecture} describes the overall architecture of an HI power delivery solution, followed by a discussion of design considerations, constraints, and goals in Section~\ref{sec:considerations}. Next, Sections~\ref{sec:analysis} and \ref{sec:optimization} discuss approaches for analyzing and optimizing these systems, discussing past work and pointing the way towards open problems and early solutions. Finally, future research directions are covered in Section~\ref{sec:outlook}, followed by concluding remarks in Section~\ref{sec:conclusion}.

\section{Power delivery architecture}
\label{sec:architecture}

\begin{figure}[t]
\centering
\includegraphics[width=\linewidth]{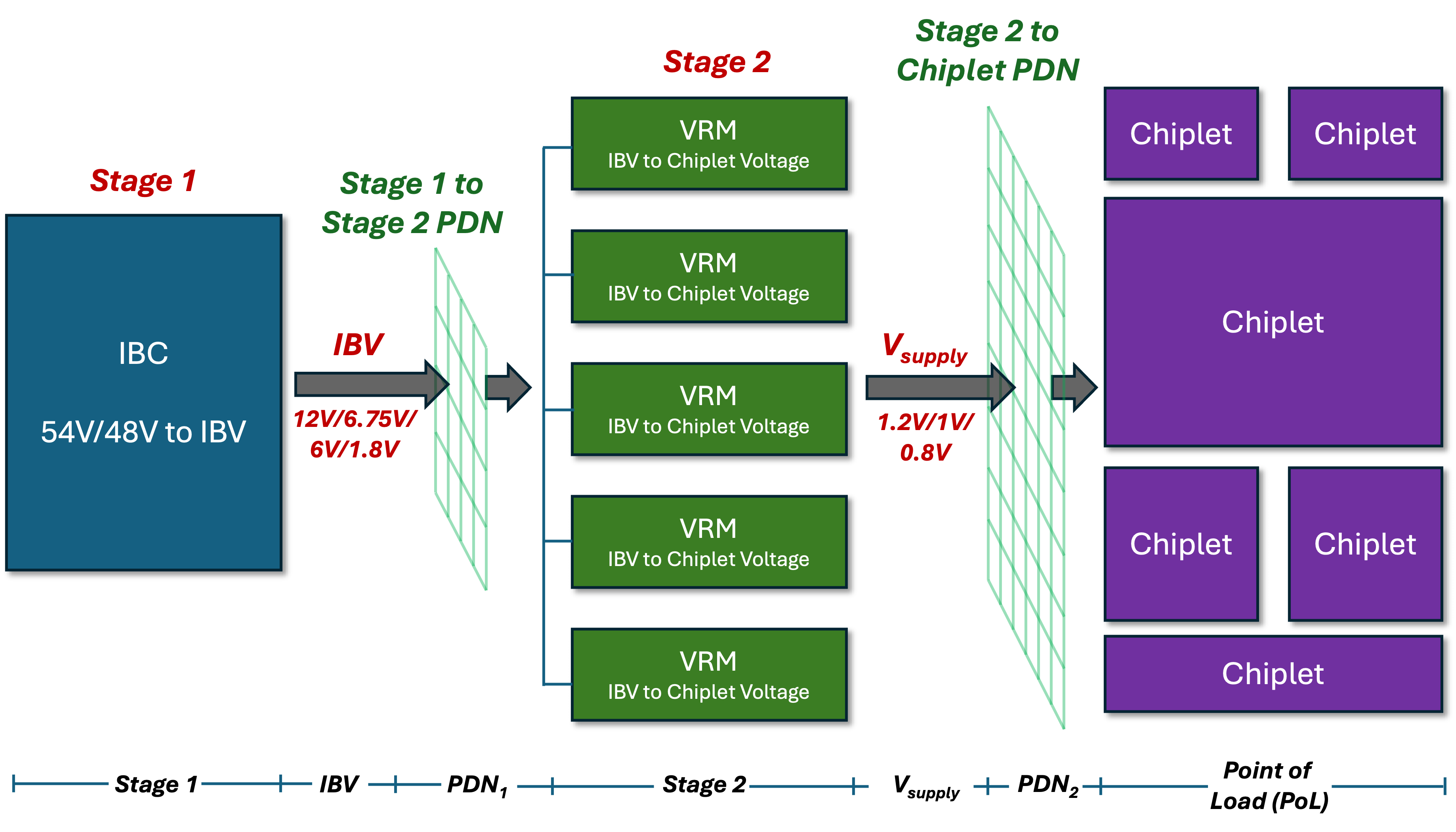}
\caption{Power delivery architecture for a 3D HI system.}
\label{fig:architecture}
\vspace{-4mm}
\end{figure}

\begin{figure}[t]
\centering
\includegraphics[width=\linewidth]{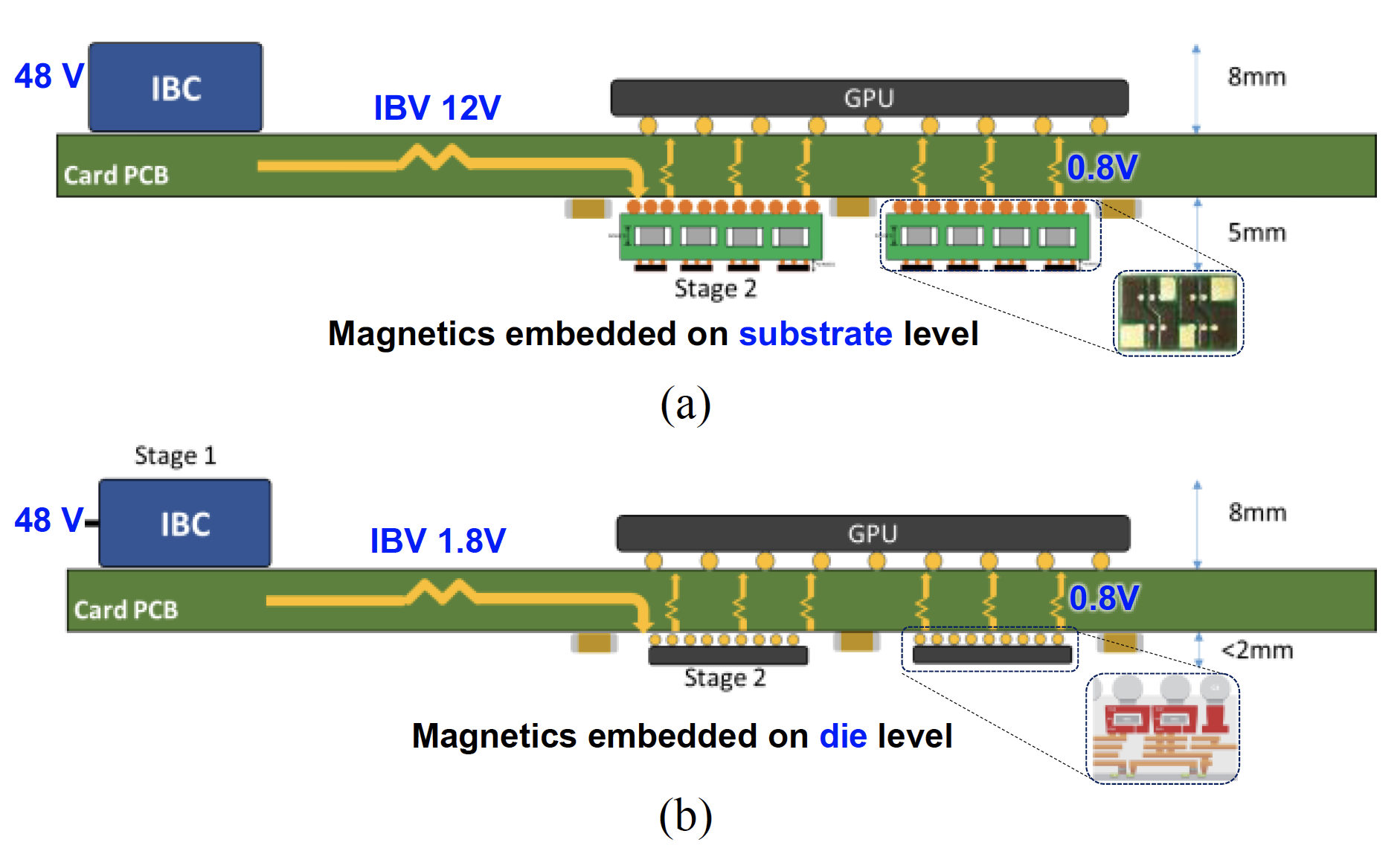}
\caption{Vertical power delivery schemes~\cite{Prakash24} with magnetics embedded at (a) the substrate and (b) the die level.}
\label{fig:Prakash}
\vspace{-4mm}
\end{figure}

\noindent
The stringent constraints on power delivery in HI systems necessitate the development of a design architecture that operates at the printed circuit board (PCB) level, the HI substrate level, and the chiplet level. The overall architecture is illustrated in Fig.~\ref{fig:architecture}, and an example of such a system is shown in Fig.~\ref{fig:Prakash}~\cite{Prakash24}. At the PCB level, off-package Stage~1 step-down power converters are used to convert high-voltage low-current power to high-current power at an IBV suitable for distribution across the package. This is then fed the package through a PDN to Stage~2 voltage regulator modules (VRMs) that convert the IBV to CMOS-compatible voltage levels, such as 0.8~V.  These regulators are placed close to the load (e.g., within the interposer) so that the high-current low-voltage path is short, and they are connected to the chiplets using a PDN in the HI substrate, delivering current to the PoL at this voltage level.  Within the chiplet, on-chip voltage regulators, such as fully integrated voltage regulators (FIVRs)~\cite{Edward_FIVR}, and low-dropout regulators (LDOs) are well-established solutions.  Approaches for advanced power delivery that have been proposed in the literature~\cite{Integrated_PDin3D_BorisVaisband,Verti_PDinHI_HanhPhuc,INTACT_Vivet_Pascal,PD_for_HPMicropro_Kaladhar,hardy_111_2023} form a good basis for building a power delivery methodology.

Historically, processor power delivery in data centers has relied on multi-stage conversion. A typical chain has been 400~V-to-48~V, 48~V-to-12~V, and 12~V-to-$\sim$1~V, with 12~V serving as the distribution bus~\cite{yeaman_datacenter_2007,Krein17,Morra25}. More recently, 48~V or 54~V DC has become the distribution voltage~~\cite{Sandri17,cong_wang_quantitative_2014,fukui_hvdc_2010,kumar_breaking_2026,lyu_composite_2018,li_google_2019,nabih_low-profile_2021,fu_10mhz_2026,winkler_increasing_2026,fang_961_2025}, and two architectural options are commonly used: (i)~two-stage conversion with a local intermediate bus voltage, or (ii)~direct single-stage conversion from 48~V to core-level voltages.

In two-stage solutions, a front end implemented with an inductive LLC, DC transformer, switched-capacitor, or hybrid switched-capacitor provides a fixed conversion ratio, followed by a regulated point-of-load multi-phase buck converter~\cite{baek_lego-pol_2020,baek_lego-pol_2019,elasser_mini-lego_2024,jiang_switched_2019,li_98.55_2018,zhu_family_2019,zhu_switching_2026,wu_hybrid_2024,zhu_500-48--1-v_2023}. Placing the buck stage close to the processor cores improves the speed of the transient response, which is critical for bursty workloads such as those for AI acceleration, with the ability to step rapidly between near-zero and full load. The achievable current-slew rate is ultimately limited by inductor dynamics; multi-phase buck converters can approach this limit by driving selected switches on/off to apply positive or negative voltage across the inductors for a full switching period and quickly ramp currents to their target values~\cite{baba_benefits_2012,parisi_multiphase_2017,zhang_wide-bandwidth_2006}. In contrast, direct single-stage conversion from 48~V-to-core voltage has been demonstrated using isolated converters as well as switched-capacitor-based hybrid topologies~\cite{zhu_dickson-squared_2022,das_regulated_2019,kumar_high-performance_2018-1,zhu_comparative_2024,ahmed_high-efficiency_2017,figueroa_2200a48v--1v_2026}. Hybrid switched-capacitor converters with multiple inductors are constrained by flying-capacitor voltage balance and typically cannot sustain full on/off operation for fast current changes~\cite{zhu_comparative_2024,das_topologies_2022}. Transformer-based isolated~\cite{kumar_high-performance_2018} or non-isolated converters~\cite{ahmed_startup_2017} often rely on frequency and/or phase-shift modulation, which generally responds more slowly than duty-cycle control and can further limit transient performance.

As a result, future data centers will likely favor two-stage architectures, with a vertically delivered point-of-load second stage placed close to the compute die. Recent demonstrations of direct conversion from the distribution bus to a local intermediate bus---bypassing the 48~V board-level bus---further reinforce this trend and push the first stage toward higher conversion ratios~\cite{navitas_navitas_2026,manners_ti_2026,noauthor_epc_2025,ahmad_transition_2025,Ye26}.

For a two-stage scheme, having chosen the input voltage level, the next degree of freedom lies in selecting the IBV.  For a 48~V / 54~V board-level input, the IBV has traditionally been chosen to be 12~V, but recent work has explored the use of lower values. For example,~\cite{Prakash24} uses a value of 1.8~V, as shown in Fig.~\ref{fig:Prakash}; \cite{Kong25} uses 1.83~V; \cite{Ahmed21} uses 6~V; and \cite{Gan24} uses 6.75~V.  Lower voltages significantly improve the efficiency of Stage~2 regulators and reduce their size.  The trade-off is that a lower IBV raises the bus current for a given power, which increases the bus I$^2$R distribution loss and reduces the Stage 1 conversion efficiency.  With an appropriate IBV, however, an overall increase in system efficiency is achievable~\cite{Ahmed21}.  An exposition of the need for this optimization for the ``last inch'' problem of power delivery, responsible for a significant fraction of losses, with first-order analyses and intuition, is presented in~\cite{Ye26}. With the advent of 2.5D and 3.5D integration, this last inch problem is being transformed to a ``last millimeter'' problem, as regulation solutions move from the motherboard into the package.

Additional gains in efficiency are achievable through the appropriate placement of the VRMs.  Traditionally, the VRMs were placed on the board, in the same plane as the HI package and adjacent to it.  This choice results in substantial routing parasitics in the supply/ground lines to the package.  More recently, designs have shifted toward vertical power delivery, placing the VRMs directly beneath the package to reduce these parasitics.  Two options are possible: the inductors are embedded in the PCB substrate (Fig.~\ref{fig:Prakash}(a)), or within the die or the HI package (Fig.~\ref{fig:Prakash}(b)).  Vertically stacked regulators have also been proposed for Stage~1 VRMs~\cite{Baek22,Vertical_PD_Inna,Krishnakumar24}.

At the chiplet level, voltage regulation can be implemented using switched-capacitor voltage regulators (SCVRs)~\cite{Le13,Harjani14} or LDOs~\cite{LDO_regulator_2014,GPU_Acceleration,LDOs_2018}. Under heavy loading and fast load transients, the large load current develops a supply droop across the finite regulator output impedance and the PDN impedance, and the I$^2$R loss in the PDN grows quadratically with that current.  In particular, the I$^2$R loss in the PDN increases rapidly as voltage droop worsens~\cite{Vertical_PD_Inna}.  Moreover, the feasible converter topology is strongly influenced by the HI substrate and its ability to integrate passives.  In principle, an HI substrate may employ capacitive or inductive converters. In a silicon interposer, where it is difficult to build sufficiently high-Q inductors, SCVRs can be used instead of inductive regulators to convert an input IBV to the output level(s) required by the chiplet.  Other types of substrates (organic or glass) can more readily support embedded magnetics and thicker conductors with lower losses, facilitating high-power inductive conversion, particularly for large-area designs.  It is well known that the use of distributed smaller voltage regulators, placed close to the load, provides significant advantages over a single large regulator~\cite{Zhou14,LDO_regulator_2014}. Selecting converters over the space of design choices, and optimally placing them to meet power integrity constraints, is a pressing issue that must be addressed in HI systems, and calls for codesign of the converters with the PDN for maximum effectiveness of the power delivery solution. Methods for automatic PDN construction, such as~\cite{Chhabria22-OpeNPDN}, may be useful when adapted to the HI context, and can be coupled with algorithms for decoupling capacitor placement within the HI substrate~\cite{KWSu03, Popovich08}.

The problem of optimizing the power distribution system involves a determination of how the losses are partitioned among various components--the Stage~1 and Stage~2 converters, and the PDNs.  This optimum depends on the sensitivity of each component: if the cost function, which may include the resource utilization (e.g., area) of the components is more prohibitive for one component than the other, the losses may be unevenly split. Generally speaking, Stage~1 converters, which operate at higher voltages, have better efficiency than Stage~2 converters, and it would be expected that the losses in the Stage 2 converters, which supply a low rail voltage at high current, may dominate those in Stage 1. Limitations on the ability to use high-Q inductors could further result in larger losses.  In other words, the loss allocation problem is an integral part of system optimization.

\section{Considerations for power delivery}
\label{sec:considerations}

\noindent
The ability of integrated circuits  to meet their functional and performance specifications is contingent on providing robust and stable power supply levels.   For example, a digital circuit is merely a practical implementation of an abstraction that models binary logic with two distinct voltages, typically the supply voltage, $V_{dd}$, and the ground potential of zero, and the gates in a circuit act as switches that steer either the $V_{dd}$ or ground signal to the output, depending on the logic level.  The ability to supply reliable $V_{dd}$ and ground levels not only dictates the correctness of computations, but also determines circuit speed~\cite{Su03}.  AMS/RF blocks also rely on well-regulated supply rails and a low-impedance ground reference to meet specifications on accuracy, linearity, and noise.

The requirements for power delivery involve the ability to deliver high-quality supply and ground voltage levels to the computing elements with high efficiency (i.e., low conversion and distribution losses), while ensuring reliability over the lifetime of the system, under resource constraints~\cite{Su03}.   Nonidealities in the power delivery path, including finite resistance, inductance, and capacitance in the lines carrying supply and ground signals, introduce voltage deviations that can shift circuit performance and timing~\cite{Jiang24}.  Today's chiplets operate at supply voltages in the range of 0.6--0.8~V, potentially going down to 0.5~V in the future, and have very little headroom for noise.  A typical noise allowance is 10\% of $V_{dd}$, of which about 2--3\% is allocated for DC noise due to currents drawn during steady-state operation. As power management strategies have grown more aggressive and multicore and multichiplet systems involve an increasing number of subsystems that may switch in and out of the power delivery system, the dominant part of the margin has been reserved to provision for transient noise.

Moreover, it may often be the case that different chiplets may operate at different supply voltage levels, or that some chiplets (e.g., analog parts) require separate and clean voltage supplies free of the noise injected by digital switching, or that chiplets may require support for dynamic voltage and frequency scaling (DVFS) whereby the supply voltage is varied according to the performance, thermal, and power constraints on the system~\cite{Rabaey09}. In such a case, there is a need to support multiple power domains.  This can be achieved through separate PDNs for each power domain driven by independent regulators, with DVFS being implemented by altering the regulator output according to the needs of the power domain.

Specific considerations in the design of the power delivery system are listed below:
\begin{description}
\item [Resource limitations] at various levels -- in connections from the package to the substrate; in wiring within the HI substrate and the chips, from the substrate to the chiplet, and within a 3D-stacked chiplet -- imply that access to the computational elements must traverse multiple bottlenecks.  Furthermore, these pins and interconnect resources must be shared with other signals, such as clock networks, chiplet-to-chiplet interconnects, and within-chip routing; with MIM and MOM capacitors; and with test infrastructure.  This is a substantial change over the already-constrained scenario that requires resource sharing in conventional 2D chips~\cite{Su:dac02}, and implies the need for in-package and on-chiplet voltage regulation. Such an approach must leverage within-package and within-substrate passives, and novel power delivery schemes, as well as coordinated mechanisms for resource sharing for limited pin and interconnect resources.
\item [Conversion efficiency losses] within multiple stages of voltage regulator modules can result in the dissipation of power before it reaches the computational elements.  Not only is this wasteful, but it results in the generation of heat flux that must be removed from the system to maintain reasonable operating temperatures.
\item [Parasitic losses] result in degradations in voltage levels, and are caused by the RLC parasitics associated within the entire power delivery stack, including package pins, redistribution layers, package wires, wiring and vias in the HI substrate, microbumps, hybrid bonds, and wires/vias within chiplets.  For example, as the current through a wire increases, its IR losses increase, or when the dI/dt through a package pin is high, the L dI/dt associated with the pin inductance can cause voltage levels to degrade.  Therefore, these losses vary during the operation of the system and are dependent on system activity.
\item [Reliability failures] are caused by aging effects induced by high currents and voltages in the system.  These include electromigration in wires, vias, and bumps, which is particularly acute in the power delivery network whose wires typically carry unidirectional currents over many years, resulting in the migration of metal over an extended period of time, as well as device failures due to phenomena such as bias temperature instability (BTI), hot carrier injection (HCI), and time-dependent dielectric breakdown (TDDB) of gate oxides, all of which are associated with some combination of high currents, high voltages, and high switching rates.  The power delivery solution must be built to last over the entire lifetime of the system in the presence of these stressing factors.
\end{description}

In the remainder of this paper, we will outline the first steps towards solving these problems and outline existing approaches that can be used as a stepping stone to building point solutions and design methodologies for power delivery.
\section{Modeling and analysis}
\label{sec:analysis}

\noindent
In this section, we address the issue of modeling and analysis of various parts of the power delivery solution.  In Section~\ref{sec:Stage1}, we describe techniques for modeling voltage regulators.  Next, we discuss techniques for analyzing the PDN: this treatment is general enough that these techniques can be used for the networks between Stage~1 and Stage~2, between Stage~2 and the chiplets, and on individual chiplets, followed by an overview of how worst-case excitations can be determined. Finally, we discuss thermal and reliability considerations in Sections~\ref{sec:thermal} and~\ref{sec:reliability}, respectively. As stated earlier, most of these solutions have not been explicitly explored in the 3D HI context, but methods from power delivery at the PCB level or the on-chip level may be adapted to this problem. The aim of our discussions in Sections~\ref{sec:analysis} and \ref{sec:optimization} is to enumerate the space of possible solutions for circuit modeling, analysis, and optimization.

\subsection{Modeling and design considerations for Stage~1}
\label{sec:Stage1}

\noindent
Power converters are modeled analytically to estimate the performance around nominal operating points and possible transients prior to hardware design. Efficient power delivery, steady-state, and transient regulation are the primary driving factors for design. Area and volume constraints for a target efficiency also play a significant role.

\begin{figure}[t]
\centering
\includegraphics[width=0.4\columnwidth]{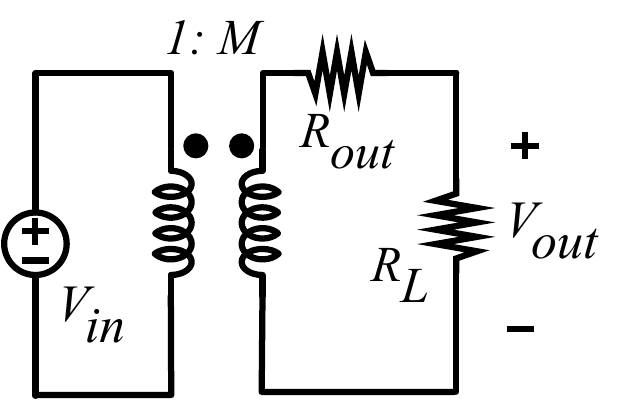}
\caption{General steady-state modeling with losses lumped into an output resistance, $R_{out}$.}
\label{fig:Steady_loss}
\vspace{-4mm}
\end{figure}

A general steady-state model for a power converter is shown in Fig. \ref{fig:Steady_loss}. This type of modeling is utilized to estimate and optimize losses around the nominal operating points of the converter ~\cite{wester_low-frequency_1973, middlebrook_continuous_1975, cuk_modelling_1978}. In this model, for a converter with an input-output relationship of $V_{out}=MV_{in}$, the losses are modeled as a lumped output resistance, $R_{out}$, and a lossless DC transformer with conversion ratio, $1:M$ (the lossless DC transformer is used only for modeling purposes).
In general, the losses of a converter depend on the operating frequency and the resistance of the current conduction paths. Operating-frequency-dependent losses are incurred by charging and discharging parasitic capacitors, hard-charging power transfer capacitors, and hysteresis and eddy current flow in magnetic cores~\cite{urling_characterizing_1989}. On the other hand, any resistance in the converter, i.e., switch on-resistance, copper trace or wire resistance and equivalent series resistance (ESR) in the capacitors and inductors, are responsible for conduction losses. These resistances also incur frequency-dependent AC conduction losses. For Stage~1, where most of the magnetic elements, inductors and transformers are implemented in PCB integrated magnetic devices, frequency-dependent conduction losses can be significant due to skin and proximity effects. The design choice for the switches, capacitors, and magnetic elements is estimated and optimized so that within the given constraints of area or volume, the lumped resistance is minimized or efficiency is maximized.

Switching power converters transition into different states depending on their switch positions and can be modeled as piecewise linear elements within one transitional state. However, a switching power converter is a nonlinear system and a linear time-invariant analytical model is used for its analysis. A widely-used method is state space analysis (SSA)~\cite{middlebrook_general_1976}, in which the state equations of the system are  perturbed, and linearized to determine the small-signal model.

\begin{figure}[t]
    \centering
    \subfloat[\label{fig:2_switch}Complementary switches]{\includegraphics[width=0.5\columnwidth]{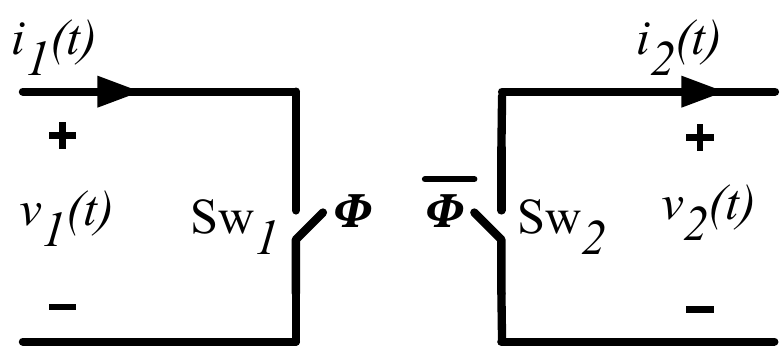}}

    \subfloat[\label{fig:2_switch_model}Small-signal model ]{\includegraphics[width=0.7\columnwidth]{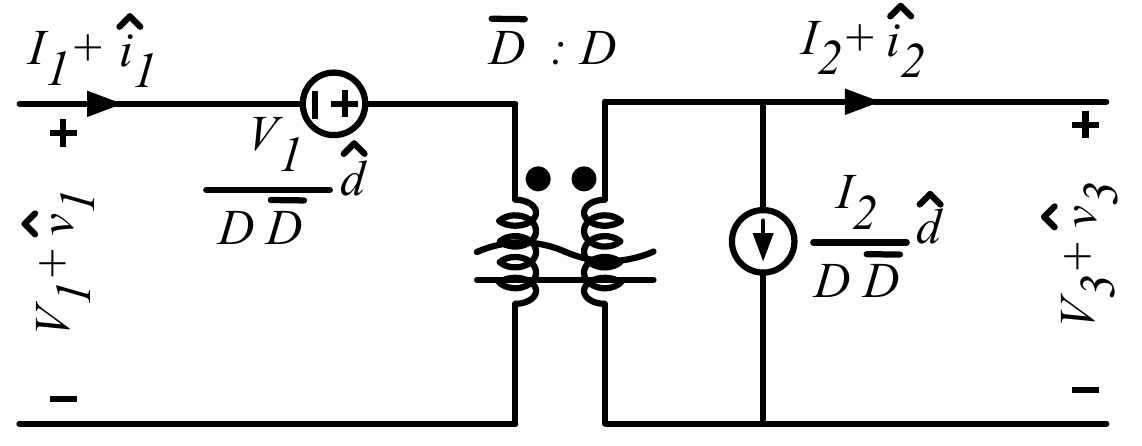}}
    
    \caption{Complementary switches and their small-signal model in traditional PWM converter, i.e., buck, boost, buck-boost, \'{C}uk, SEPIC, flyback, etc.}
    \label{fig:2_switch_and_model}
    \vspace{-4mm}
\end{figure}

An alternative approach to SSA is averaged switch modeling (ASM). In this method, complementary switches in a power converter are identified and their voltage and current waveforms are analyzed~\cite{wester_low-frequency_1973, vorperian_simplified_1990-1}. The waveforms are directly averaged and then perturbed and linearized to determine the DC operating points and AC response. Finally, complementary switches are replaced with small-signal perturbation sources from linearization, a transformer with AC and DC transfer capability. In Fig.~\ref{fig:2_switch_and_model}(a), two complementary switches with voltage and current stresses that vary over time are shown. After averaging, perturbation, and linearization, their small-signal analytical model is determined as in Fig.~\ref{fig:2_switch_and_model}(b).

\begin{figure*}[t]
\begin{minipage}{1\textwidth}
    \hfill{}
    \subfloat[\label{fig:SCBC2}2LSCBC]{\includegraphics[width=.3\textwidth]{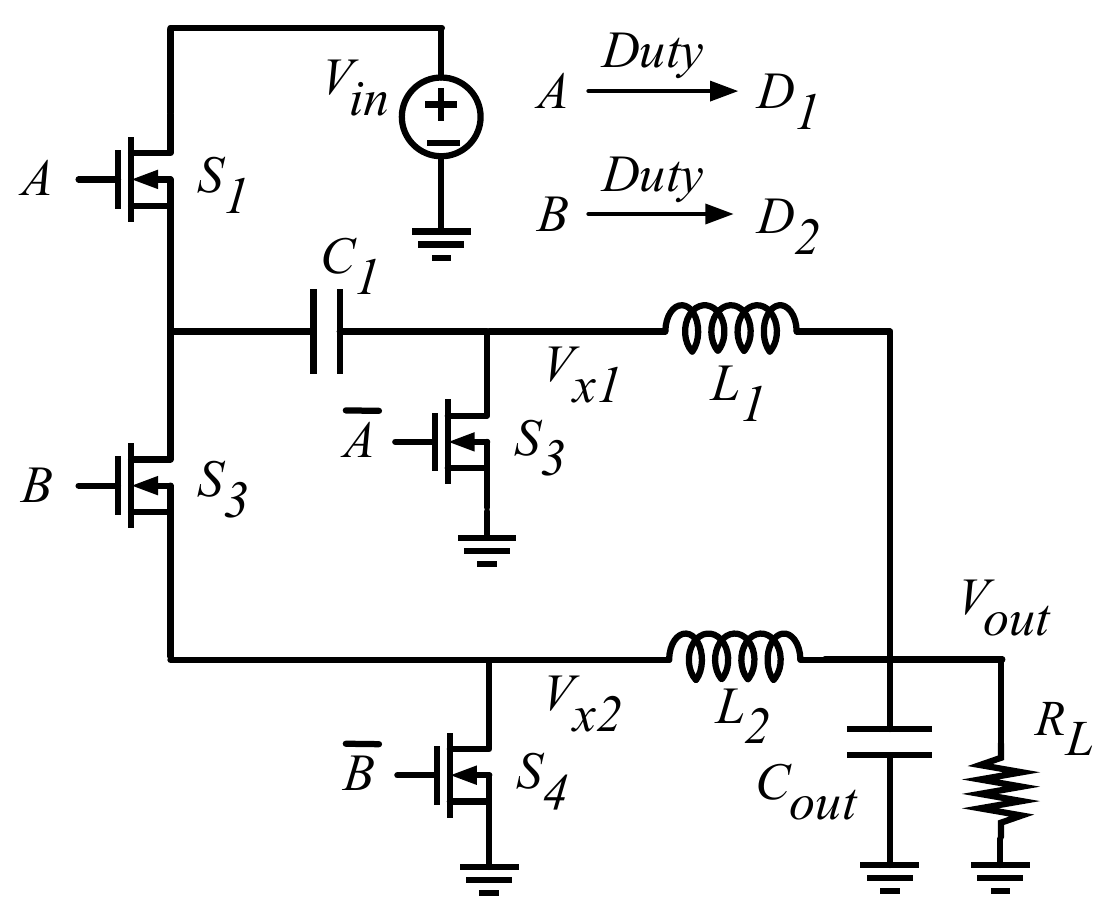}}
    \hfill{}
    \subfloat[\label{fig:SmallSCBC2}Small-signal model of 2LSCBC]{\includegraphics[width=0.5\textwidth]{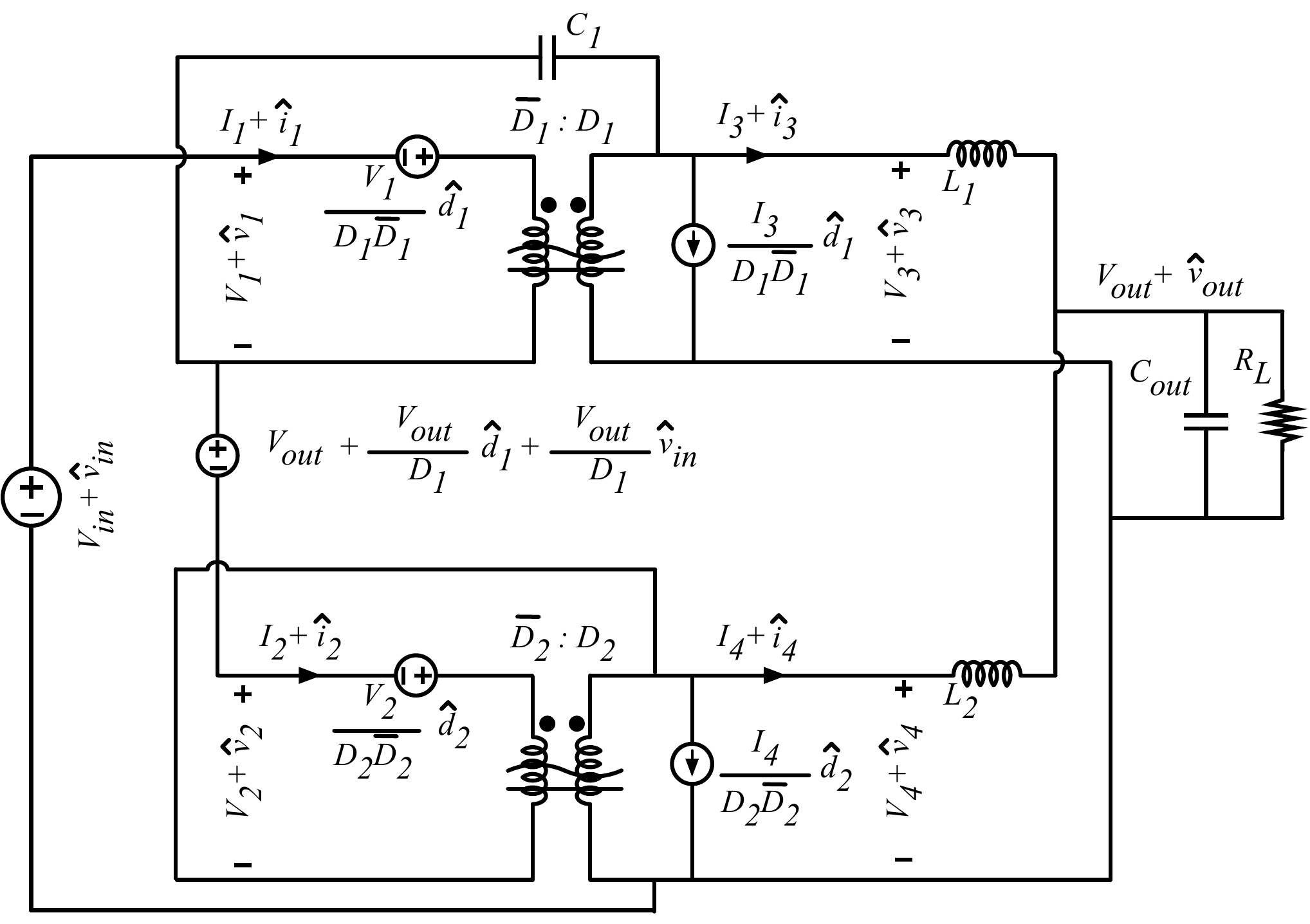}}
    \hfill{}
    \caption{(a)~Schematic of a two-level series capacitor buck converter (2LSCBC) and (b)~its small-signal model.}
    \label{fig:SCBC2andsmall}
    \vspace{-12pt}    
\end{minipage}
\end{figure*}

Note that SSA and ASM are low-frequency approximations of sampled-data modeling \cite{brown_sampled-data_1981,lau_small-signal_1986}. However, they remain relevant to date due to their simplicity. Both result in the same small-signal models: SSA is more equation-focused, while ASM is circuit-oriented. Both methods generate models with DC operating points and small signal AC characteristics. The SSA method, which is widely used by practitioners, is solved by averaging the state equations of a power converter. Even for known power converter topologies, the state equations must be solved again to obtain the small signal equivalent circuit. In contrast, ASM involves averaging the voltage and current waveforms only for the switches in the converter. Overall, ASM is computationally superior to SSA while providing same results; moreover, it can be easily extended to new and unknown converter topologies by inspection, unlike SSA which lacks this advantage for quick and reliable design.

Parasitic resistances and capacitances can also be included in the small-signal model in Fig.~\ref{fig:2_switch_and_model}(b), and their effects can be analyzed for both the AC response and the DC performance. Including parasitic capacitances and/or resistances (i.e., ESR of the output capacitor) may result in a higher order system or new converter dynamics, which will have effects for practical engineering considerations. The extra-element theorem can be applied to get the new small signal models directly from the already-derived ideal small signal models~\cite{middlebrook_null_1989}. Including parasitics can change the DC operating points of the converters. The simplification of small-signal models from ASM, ignoring the DC components, results in a model similar to Fig.~\ref{fig:Steady_loss}.

Applications of ASM can be extended to complex Stage~1 converters, such as the widely-used multilevel series capacitor buck converter (SCBC)~\cite{shenoy_comparison_2016,das_regulated_2019, elasser_mini-lego_2023}. The schematic of a two-level SCBC (2LSCBC) is presented in Fig.~\ref{fig:SCBC2andsmall}(a). The small-signal model of this converter is shown in Fig.~\ref{fig:SCBC2andsmall}(b), derived from the ASM method~\cite{das_averaged_2023}. The small-signal models of multilevel SCBC converters coincide with conventional multiphase buck converters~\cite{das_averaged_2023,zhang_wide-bandwidth_2006}. Other types of converters used in Stage~1~\cite{kumar_high-performance_2018-1,ahmed_single-stage_2020}, can also be modeled with ASM or SSA.

\begin{figure}[t]
\hfill
\includegraphics[width=1\columnwidth]{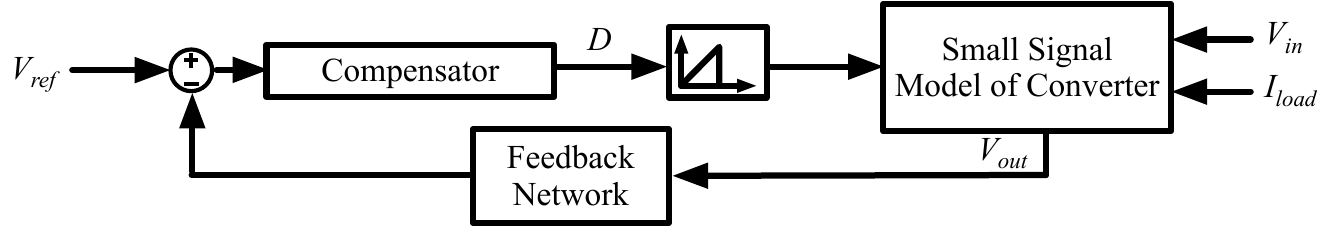}
\hfill\hfill
\caption{Block diagram of Pulse Width Modulation (PWM) control method for a DC-DC converter.}
\label{fig:Small}
\vspace{-4mm}
\end{figure}

To analyze the transient performance of the full system, the small-signal models are assembled as shown in Fig.~\ref{fig:Small}, which shows the block diagram of a pulse width modulated (PWM) control for a converter. The ASM techniques and results can also be modified for different modes of operation and for other types of control~\cite{vorperian_equivalent_1989,vorperian_simplified_1990-1,vorperian_simplified_1990-2}. The overall performance of the converter for steady-state regulation and transient recovery time, overshoot, and undershoot can be estimated using the averaged circuit model and the control block diagram.

Most recently explored Stage~1 power delivery topologies are based on switched-capacitor (SC) hybrid converters~\cite{baek_lego-pol_2020,baek_lego-pol_2019,elasser_mini-lego_2024,jiang_switched_2019,li_98.55_2018,zhu_family_2019,zhu_switching_2026,wu_hybrid_2024,zhu_500-48--1-v_2023}. However, several SC-based hybrid configurations exhibit voltage imbalance in the power-transfer flying capacitors. Prior work has investigated the origins of these imbalance mechanisms and identified key contributing factors~\cite{rentmeister_48v:2v_2017,stillwell_active_2019,xia_state_2019,celikovic_modeling_2019,das_demystifying_2019}.

It has also been shown that coupled inductors can introduce a natural balancing effect on the flying capacitors~\cite{zhou_balancing_2024}. At the same time, the existence of balanced converter topologies that do not rely on magnetic coupling indicates that capacitor voltage balance is governed by more fundamental circuit properties, independent of coupled inductors. Despite this, coupled inductors have become a prevalent design element in PWM-operated hybrid converters. Although modeling methods for coupled inductors have been developed~\cite{wang_matrix_2022}, their broader impact on hybrid converter dynamics and performance remains insufficiently understood and warrants deeper investigation.

In addition, high-current applications typically employ multiple converters connected in input-parallel/output-parallel configurations to distribute the load~\cite{baba_benefits_2012}. Various current-sharing strategies, including droop control, have been proposed to ensure uniform load distribution~\cite{qi_comparative_2024}. However, these methods fundamentally depend on accurate current sensing; without it, effective load sharing cannot be achieved~\cite{rincon-mora_switched_2023}. This highlights the need for improved lossless current sensing techniques that can operate reliably in systems with very high load currents.

\subsection{Electrical analysis of PDNs}

\noindent
The PDN is typically modeled as an RLC system that drives time-varying current sources at the loads.  The wires
in the $V_{dd}$ grid can be modeled as a linear RLC network that is connected through pads to multiple $V_{dd}$ sources, and to switching elements in the circuit that draw current through the power grid; the ground grid is analyzed analogously.  The simulation of this network requires the solution of the following system of differential equations, which may be formulated using Modified Nodal Analysis (MNA) \cite{Ho75}:
\begin{equation} \label{eqn:0a}
G {\bf x}(t) + C {\bf x'}(t) = {\bf b}(t),
\end{equation}
where $G$ is a conductance matrix, $C$ is the admittance matrix resulting from capacitive and inductive elements, {\bf x}(t) is the time-varying vector of voltages at the nodes, and currents through inductors and voltage sources, and {\bf b}(t) is the vector of independent and potentially time-varying excitations.  Under the restriction of rectilinear routing, the number of neighbors of each node is seen to be limited by a small constant, typically corresponding to nearest-neighbor nodes along the $x$, $y$, and $z$ axes, and therefore a system with $n$ nodes has $O(n)$ circuit elements. As a result, the number of nonzeros in the $G$ and $C$ matrices are $O(n)$, i.e., both matrices are very sparse.  This system of differential equations is efficiently solved in the time domain by reducing it to a linear algebraic system,
\begin{equation} \label{eqn:0b}
(G + C/h) {\bf x}(t) = {\bf b}(t) + C/h {\bf x}(t-h),
\end{equation}
using the Backward Euler (BE) technique with a small fixed time step, $h$.  Even for 2D chiplets, the size of this problem is very large, resulting in the formulation of a system of billions of equations in an equal number of variables.  When {\bf x} consists only of node voltages, as in the case of a network with only resistors, capacitors, and current sources, the coefficient matrix can be shown to be symmetric and positive definite. A symmetric positive definite formulation is feasible even when inductive elements defined by inductance matrices and independent voltage sources are included in the analysis, although this would involve an additional reduction step from the modified nodal formulation to the nodal formulation, wherein the variables for inductor currents and voltage source currents are eliminated.  On the other hand, the use of a $K$ matrix (inverse inductance matrix)~\cite{devgan:iccad00} formulation directly leads to a symmetric positive definite matrix. The positive definiteness property can be exploited by using efficient techniques such as Cholesky factorization (a direct method) or the conjugate gradient method (an iterative method), and the efficiency can be further improved by taking advantage of sparsity.

While such dynamical systems are often solved using variable time steps, in the context of power grid solution, keeping $h$ constant keeps the left-hand side matrix unchanged over time, so that the cost of Cholesky factorization of this matrix is amortized by its reuse in backward/forward substitution over a large number of time steps~\cite{Matrix_Comp_Golub_2013}. 

\begin{figure}[t]
\centering
\subfloat[]{\includegraphics[width=0.8\linewidth]{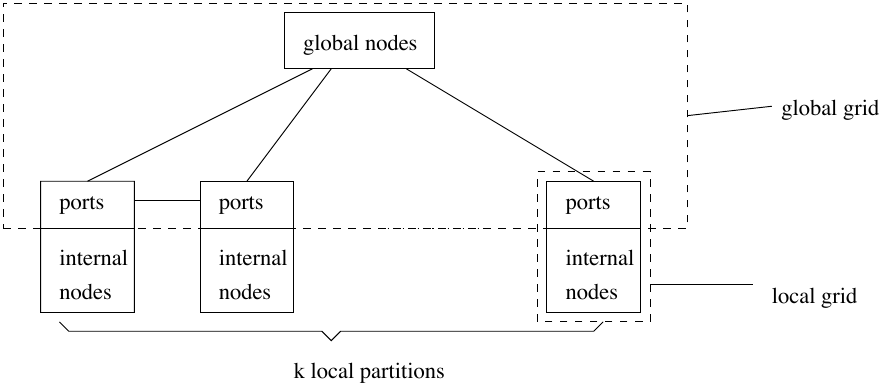}}

\subfloat[]{\includegraphics[width=0.8\linewidth]{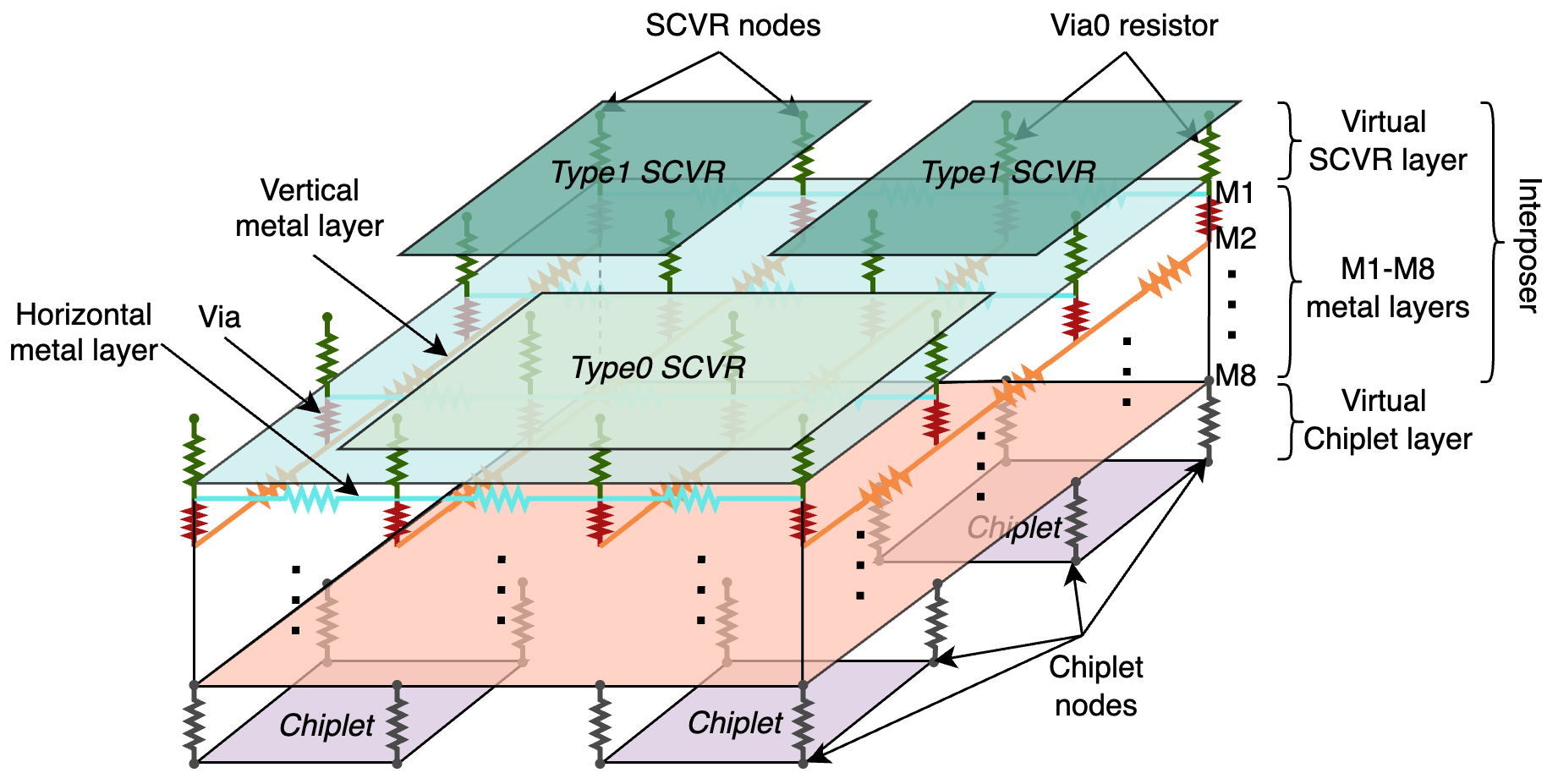}}
\caption{(a)~The natural hierarchy in the PDN, where local block grids connect to a global grid, has been used to efficiently analyze 2D chips by macromodeling local grids in~\cite{Zhao02}. (b)~Structure of a HI stack that shows SCVRs on an active interposer, connected to chiplets through eight layers of metal: while optimizing regulator locations, similar principles as~\cite{Zhao02} are used to macromodel the metal stack in~\cite{Zhang25}.}
\label{fig:hierarchy}
\vspace{-4mm}
\end{figure}

Several classes of techniques have been proposed to manage the cost of solving such large systems.  First, the natural sparsity of the system, described above, makes the solution cost only slightly superlinear, and well below the $O(n^3)$ cost of solving a general system of $n$ linear equations.  However, a major bottleneck is often not the computation, but the memory overhead associated with solving such large systems: decomposition and reduction techniques can be very effective in limiting this issue.  Since integrated systems are typically built hierarchically, constructing building blocks with their own local power grids that are then hooked up to a higher-level global power grid that connects multiple blocks, as illustrated in Fig.~\ref{fig:hierarchy}. One class of methods, exemplified by the approach in~\cite{Zhao02}, exploits this hierarchy by using sparsified Schur complements for the power grids at lower levels of this hierarchy, and successively solving smaller, and yet sparse, systems.  In~\cite{Zhang25}, a similar approach is employed to macromodel intermediate layers of the PDN during a regulator placement optimization, eliminating these nodes by creating the corresponding Schur complement. A second class of methods uses multigrid techniques~\cite{Kozhaya02,Su03b,Zhuo08}, including GPU implementations~\cite{Feng08,Feng11} that abstract away internal nodes, effectively introducing a hierarchy, to reduce the size of the system.  Successive solutions of the multigrid V-cycle~\cite{Trottenberg2000} provide a computationally efficient and accurate solution.  

\begin{figure}[t]
\centering
\subfloat[]{\includegraphics[width=0.8\linewidth]{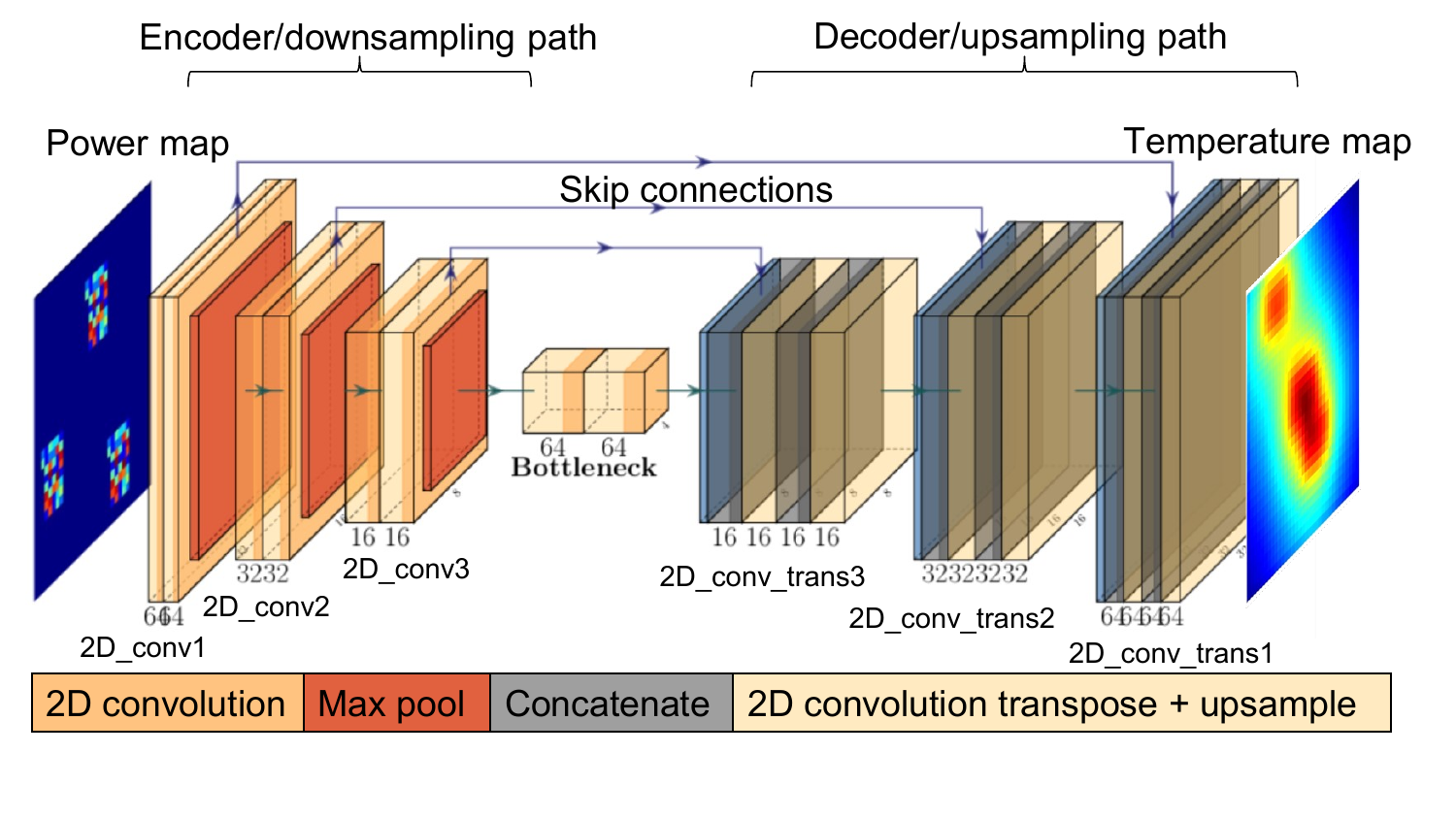}}

\vspace{-4mm}
\subfloat[]{\includegraphics[width=0.35\linewidth]{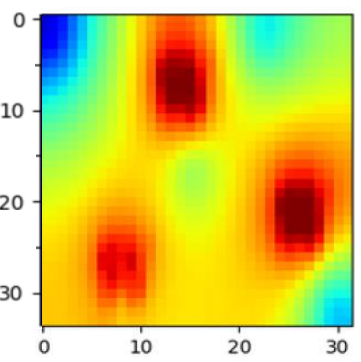}}
\hspace{4mm}
\subfloat[]{\includegraphics[width=0.36\linewidth]{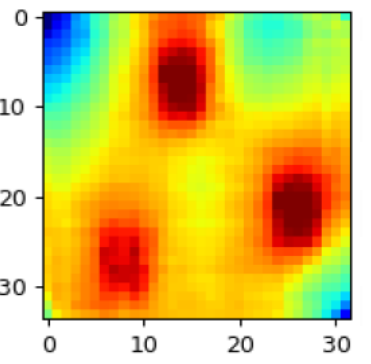}
\raisebox{2.6mm}{\includegraphics[width=0.138\linewidth]{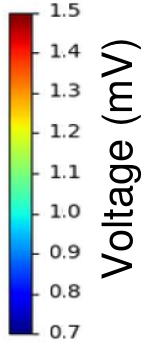}}}

\caption{Machine learning approaches for fast PDN analysis using U-Net. The topology of a U-Net is shown in (a), and the comparison of the ML prediction and exact analysis are shown in (b) and (c), respectively~\cite{Chhabria20}.}
\label{fig:mlpdn}
\vspace{-4mm}
\end{figure}

Other classes of methods solve these problems differently, without Cholesky factorization.  The efforts in~\cite{Su00,Cao02} use model order reduction techniques for the hierarchical analysis of PDNs. Most recently, various machine learning methods have been proposed to accelerate the analysis of PDNs using encoder-decoder networks~\cite{Chhabria20,Chhabria22-TODAES,Chhabria21}, as illustrated in Fig.~\ref{fig:mlpdn}, and extended to using graph neural networks~\cite{Yang25a}.

\subsection{Determining worst-case excitations}

\noindent
An open problem over many decades has been that of determining the worst-case current patterns that result in the largest voltage drop, or the worst thermal profile.  Particularly for modern AI workloads that exhibit highly dynamic, microsecond- or submicrosecond-level dI/dt transients, it is important to model and extract these high-frequency transient noises across the package-to-chip PDN interfaces in order to provide realistic excitations applied to the system.

This is a difficult problem. First, there may be no single worst-case excitation, as different excitations result in worst-case behavior in different parts of the chip.  Second, the worst case typically is a time-varying excitation rather than a static DC draw, complicating the problem by adding both a spatial and a temporal dimension.  For the on-chip case, this has been addressed using the so-called vectorless and vectored approaches.  A vectorless method attempts to identify a worst-case, often DC, current excitation that is an envelope on the average switching activity of a block. Due to its approximations, it is often pessimistic and unable to account for time-varying effects, but it is appropriate in the early stages of design.  A vectored approach, on the other hand, is based on actual traces of input stimuli based on specific operating modes of the system. Every excitation in such an approach is a real stimulus, but it depends on the ability to extract these stimuli, and the search may not be exhaustive enough to uncover the genuine worst case.  Practically, a mix of vectorless and vectored methods is used for 2D chip design, and it is anticipated that a similar approach will be used for 3D HI systems.

\subsection{Thermal considerations}
\label{sec:thermal}

\noindent
Thermal management and power delivery are two sides of the same coin~\cite{Sapatnekar09}: the power delivery scheme transmits energy into the system so that it can be dissipated during computation, but the resulting heat must be removed from the system lest it should excessively raise on-chip temperature levels.  

Today, circuits are typically limited to a peak temperature of 125C, 105C, or 85C for automotive/military, high-end consumer, and standard commercial applications, respectively. The increased level of power generation will make it challenging to stay within this envelope, and there is a significant body of work on developing thermal solutions for 3D HI systems. In the context of power delivery, it should be noted that a vast  majority of the heat in 3D HI systems will be generated by the compute units. This heat flux in itself will create thermal hot spots that affect power delivery circuits. In addition, even at 90\% efficiency, delivering 2kW of power will result in losses of over 200~W. Especially in Stage 2 converters, this dissipated power create localized hot spots within small regions, resulting in elevated temperatures that may affect the performance of power delivery circuitry.

\begin{figure}[t]
\centering
\subfloat[]{
\includegraphics[width=0.7\linewidth]{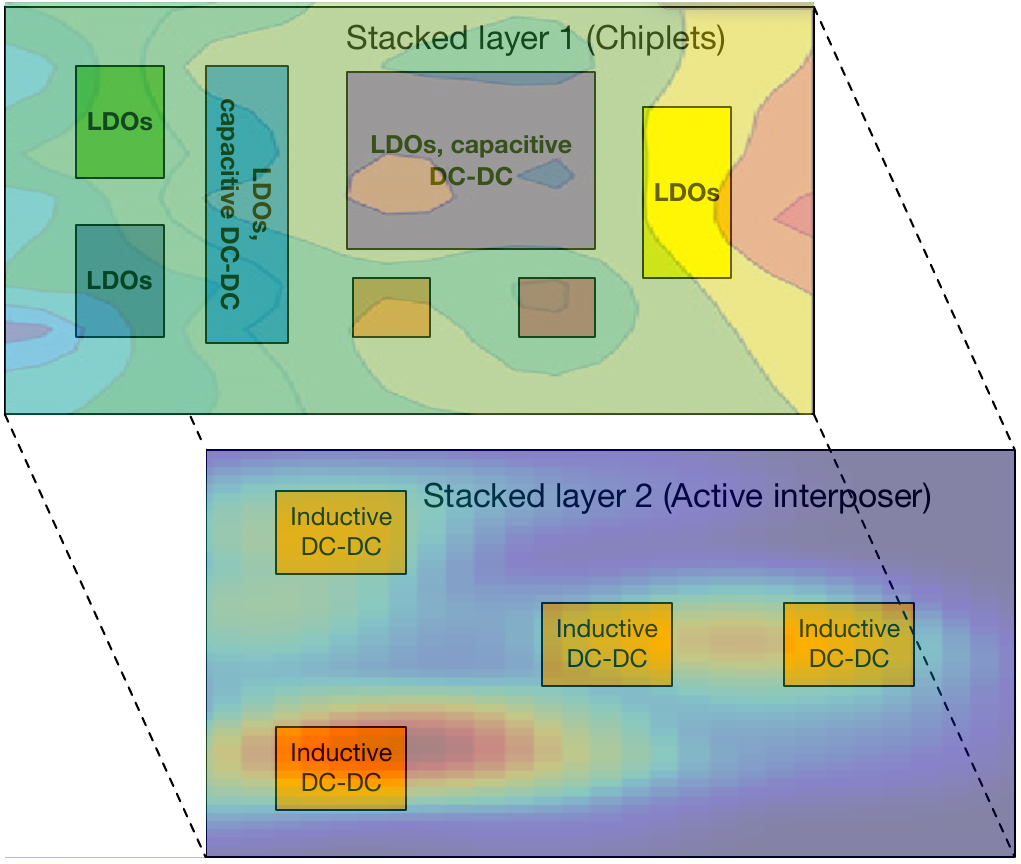}
}
\caption{Thermal contours of a 3D HI solution, illustrating how thermal contours can impact power delivery units, from on-chiplet regulators to VRMs in the HI stack.}
\label{fig:thermal}
\vspace{-4mm}
\end{figure}

From the point of view of designing elements of the power delivery solution, elevated temperatures not only cause performance degradation (e.g., by altering the carrier mobility in transistors), but can also result in accelerated aging~\cite{Zhan08}. This calls for thermal analysis techniques that can be used to identify hot-spots and support electrothermal and reliability analysis, and using them in conjunction with physical design and real-time scheduling to limit the thermal exposure of regulators and interconnects.  An example thermal map is shown in Fig.~\ref{fig:thermal}.  Note that regulators and interconnects can be heated due to (a)~the heat that they produce themselves, within transistors in regulators and due to self-heating in interconnects, and (b)~more importantly, the heat that is produced by neighboring blocks whose voltage is regulated by these elements.  This issue is particularly acute because the circuit blocks with the highest power densities have high current draws, and therefore are likely to be hotter.  However, they also require regulators to be placed close to them to supply their high current demands, making them susceptible to thermal- and reliability-induced degradations.

ML-based approaches that have shown promise for thermal analysis at the full-chip level can be extended to HI systems. For example, in~\cite{Chhabria22-TODAES}, the thermal problem is viewed as translating an image of an input power map, with knowledge of the heat removal paths, to an output thermal map using a U-Net-based approach.  In HI systems, the thermal solution may be more complex, e.g., involving cold plates or active microchannel-based cooling within the HI stack.  Such fast models can be used to drive electrothermal analysis, as has been demonstrated at the chip level for a power amplifier structure~\cite{Karmokar24}. Thermal optimization involves the use of a mix of passive and active cooling strategies. Passive cooling involves the use of thermal vias~\cite{Goplen05}, while active cooling may involve cold plates~\cite{Lorenzini16,Hoang22} near the package, or microchannels within the package~\cite{Yan25,Choi25,Choi26}.

\subsection{Electrothermomechanical reliability}
\label{sec:reliability}

\noindent
Most device reliability mechanisms are impacted by higher temperatures~\cite {Fang12}, particularly BTI and HCI, which degrade the threshold voltage and drive current of a transistor over time.  Hence, reliability must be a first-class objective during HI system design, using cooling solutions to mitigate reliability hot spots.  Electromigration (EM) is also significantly accelerated by temperature, and therefore it is essential to build thermally-aware solutions that account for EM both in wires~\cite{Najm20,Najm21,Shohel21a,Shohel21b,Shohel23,Shohel25,Mishra13} that distribute power and signals, and in bumps~\cite{Xiong14,Shen23} that connect the chiplets, substrate, package, and PCB.

Mechanical reliability also plays a very strong role in HI systems in several ways. First, substrate warping due to mismatches in the coefficients of thermal expansion (CTEs) can cause a large and thin substrate to warp.  Today's techniques capture these effects using first-order models \ignore{~\cite{Suhir09}} or detailed numerical simulations, but it will be essential to develop methods that present tradeoffs between the simplicity of the former and the accuracy of the latter, possibly driven by ML models.  Second, issues related to CTE mismatches between bumps/bonding pads and the chiplets, or TSVs and silicon, are known to cause circuit performance drifts~\cite{Marella15}: with the increasing integration of sensitive AMS/RF blocks, such analyses will become important.  Third, stress migration~\cite{Kteyan21} due to residual stress in the chip structure can exacerbate the impact of EM, and must be factored in during EM analyses.  

\section{Optimization}
\label{sec:optimization}

\noindent
In this section, we list several techniques that may be used to optimize the power delivery system.  In Section~\ref{sec:disaggregation}, we discuss the considerations for power delivery during the chiplet selection step of disaggregation, followed by considerations for optimal IBV selection in Section~\ref{sec:IBV_selection}. Next, in Sections~\ref{sec:VR_opt} and \ref{sec:PDN_opt}, we overview techniques for optimizing the regulators and the PDN, respectively.  Due to the need for close proximity between the regulators and the PoL to reduce the impact of parastics, we discuss the optimal placement of voltage regulators in Section~\ref{sec:VR_placement}.  Decoupling capacitors can be a vital tool in the quest to reduce transient noise, and HI offers new opportunities in the form of embedded passive capacitor elements: we will address these issues in Section~\ref{sec:decap}.  Finally, we present an unconventional approach for power delivery, using a multistory scheme, in Section~\ref{sec:MSPD}.

\subsection{Chiplet selection and cooptimization}
\label{sec:disaggregation}

\noindent
As an HI system is assembled using a set of chiplets placed on a substrate, various elements go into building a cost function for optimization. In addition to optimizing yield and cost, reusability can be maximized and carbon footprint reduced~\cite{Li22,Nalla25,Sudarshan24} by disaggregating the system into a set of chiplets drawn from a library of predesigned chiplets. Elements in this library could be very diverse -- e.g., built using different technology nodes, or using 2D or 3D assembly -- and could have very different power networks within the chiplet.  For example, some chiplets may have only PDNs without on-chip regulation; some also have LDOs; others may have on-chip FIVRs; yet others may be 3D-integrated with back-side interconnect for power delivery~\cite{Lin22,Veloso22,Hafez23} and TSVs across 3D tiers.  The disaggregation step that selects chiplets from the library must therefore be conscious of these capabilities and incorporate them in the cost function.  Once the chiplets are selected, the power delivery solution on the HI substrate should account for these on-chiplet regulation capabilities during optimization.

\subsection{Efficiency optimization and IBV selection}
\label{sec:IBV_selection}

\noindent
The current load in an HI system can change over time, leading to regulator inefficiency as the regulators may be operating far from its peak efficiency points.  This issues has been addressed in~\cite{Krishnakumar26ectc}, which considers the impact of dynamic power management for distributed vertical power delivery by dynamically turning some VRs on or off to ensure that the power delivery solution operates at high efficiency.

A second knob for efficiency optimization, as indicated in Section~\ref{sec:architecture}, is the IBV, which sits at the center of the power delivery tradeoff as it couples the bus I$^2$R loss to the losses of the first and second stage converters. This dependency carries over into the PDN, where the voltage seen by each chiplet is shaped by where the second stage converters are placed and by how the current spreads through the network. The picture is complicated further by DVFS, under which the supply voltage, the load current, and the switching activity all change from one operating state to the next.  Moreover, an IBV that is efficient at one operating point can become inefficient, or even infeasible, at another. A number of designs have chosen particular values, weighing 12~V against lower options in board-level systems, adopting 6~V and 12~V in data center platforms~\cite{Ahmed21}, and using 1.8~V in vertical implementations~\cite{Prakash24}. These efforts confirm that the IBV strongly determines the efficiency, yet each stops short in one respect: some fix the IBV by rule of thumb rather than by formal optimization, others evaluate it at a single operating point and overlook the effect of DVFS, and still others treat the power stages in isolation and miss the coupling between them.  A method that considers the impact of fast ns or sub-ns power management, which requires faster capacitive voltage converters, has been proposed in~\cite{Yue26}.  Scaling such methods to high power levels and fast response times remains an open problem.

\subsection{Regulator selection and optimization}
\label{sec:VR_opt}

\noindent
At each stage, the regulator topology and its design point are a key degree of freedom, set by the input voltage, the substrate technology, and the power demand of each chiplet.  The first stage inductive regulator bridges the high voltage gap (e.g., 48~V/54~V to IBV). Given this high input voltage, it is typically implemented in a high-voltage technology such as gallium nitride (GaN) or bipolar-CMOS-DMOS (BCD), using the modeling techniques outlined in Section~\ref{sec:Stage1}.  When the IBV is sufficiently low, the second-stage converter can be built using standard CMOS technologies. While inductive converters are commonly used at the board level in HI systems, inductive conversion can often be avoided in silicon-interposer designs for the downstream distribution of bus voltages to the chiplets, and integrated voltage regulators, in particular high-efficiency SCVRs~\cite{Le13,Harjani14}, are attractive for their small size, design flexibility, and high efficiency. To serve the multiple output levels a chiplet may require, a reconfigurable SCVR switches among several topologies, each realizing a discrete conversion ratio, and obtains the intermediate voltages by regulating its output impedance~\cite{Hanh_DesignTech_of_SCVR}.  

Prior work has investigated modeling and optimization techniques for individual regulators -- SCVRs~\cite{Hanh_DesignTech_of_SCVR, Seeman_Analysis_optimze_SCVRs, Converter_sizing_Julien, Anderson_pareto_opt_of_SCVRs}, LDOs~\cite{LDO_regulator_2014,GPU_Acceleration,LDOs_2018}, and FIVRs~\cite{Edward_FIVR}: due to space limitations, these are not described here.  Moreover, the physical location of the regulator is itself a part of this selection: an inductive second stage is typically realized directly below the HI stack as in Fig.~\ref{fig:Prakash}, whereas a switched-capacitor second stage can be embedded within the interposer~\cite{INTACT_Vivet_Pascal}, a choice that follows from the regulator technology and the passives available in the substrate.

Each of these regulators has been optimized as a standalone block. What HI lacks is a methodology that selects and designs them jointly with the IBV, the substrate passives, and the heterogeneous chiplet demands, so that the regulators are optimized as part of the whole power delivery system rather than in isolation.

\subsection{PDN optimization}
\label{sec:PDN_opt}

\begin{figure}[t]
\centering
\subfloat[]{
\includegraphics[width=0.4\linewidth]{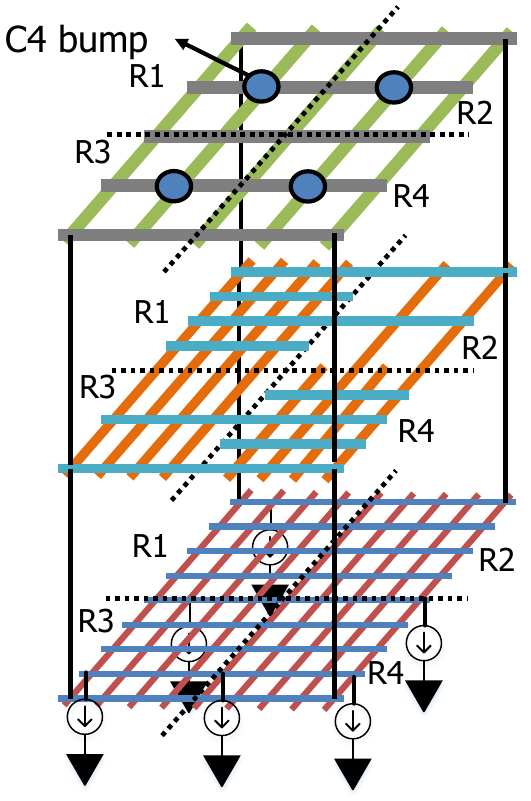}
}

\subfloat[]{
\includegraphics[width=0.5\linewidth]{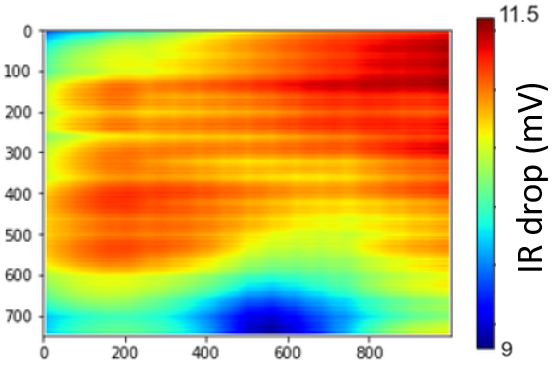}
}
\subfloat[]{
\includegraphics[width=0.475\linewidth]{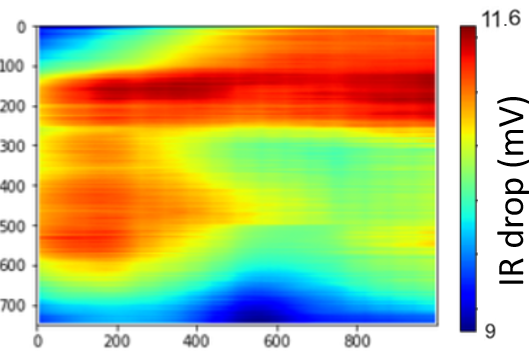}
}
\caption{(a) A template-based PDN with piecewise-uniform pitches, and the resulting IR drop maps after~(b)~early-stage and (c)~late-stage design.}
\label{fig:pdn}
\vspace{-4mm}
\end{figure}

\noindent
In the on-chip context, the problem of PDN optimization has been extensively studied, and these methods are applicable to optimizing the PDN topology of 3D HI systems.  These methods generally use approaches such as topology selection and wire widening~\cite{Su00,Tan03,Singh05,Todri14} in order to reduce the effective resistance of the grid.  The need for fast design-space exploration motivates the use of machine learning methods for this problem. In \cite{Chhabria22-OpeNPDN}, an automated machine-learning-based approach for developing a PDN topology using predefined template structures was proposed. An example of a PDN built using a set of templates is shown in Fig.~\ref{fig:pdn}(a): the layout area is tessellated into regions such that each region can use one of the available templates. Each template spans from the top to the bottom metal layer, and the template pitches are chosen so that the templates can be freely abutted to maintain connectivity of the overall PDN. In an early design stage, with coarse estimates of current, congestion, blockages, and pin locations, an initial grid is synthesized. As the design is better defined at a later stage, and fine-grained distributions of current and congestion become available, the PDN template choices are incrementally refined. At each stage, a trained classifier, based on a convolutional neural network (CNN), selects the optimal template. In late-stage design, an additional input biases the classifier toward an option with an option that is close to the early-stage choice. The IR drop maps at the end of early-stage and late-stage optimization, computed under a static current load, are shown in Figs.~\ref{fig:pdn}(b) and (c), respectively, and can be seen to be self-consistent. In principle, this methodology can be extended to the design of PDNs within the HI substrate, matched with on-chiplet PDNs to deliver reliable supply levels. Extending this methodology to PDNs within the HI substrate, matched to the on-chiplet PDNs, additionally requires transient-aware optimization and co-design with regulator and decoupling-capacitor placement, since the static-IR-drop formulation alone does not capture the dynamic and cross-layer effects that dominate HI power integrity.

\subsection{Regulator placement}
\label{sec:VR_placement}

\noindent
In HI systems, almost every performance parameter is tightly coupled to physical design. Decisions that are made during layout impact the spatiotemporal distribution of power dissipation in the system, and hence the power delivery solution and the thermal map of the system, which in turn affects the performance and reliability of elements of the power delivery solution, such as regulators and PDN interconnects. Therefore, placement considerations must incorporate the topology of the power delivery solution.  This can be achieved by developing PDNs methodologies as part of physical planning; by using distributed regulators, often more efficient than centralized regulators; by interleaving the placement of regulators, decaps, and inductive elements within the array of chiplets, or within the substrate when possible, to integrate chiplet placement with the optimization of the power delivery solution; and by managing the sharing of interconnect resources between PDNs and signal/clock nets. Some of these issues have been addressed at the chip(let) level in the past.

\begin{figure}[t]
  \centering
  \subfloat[]{\includegraphics[width=0.49\linewidth]{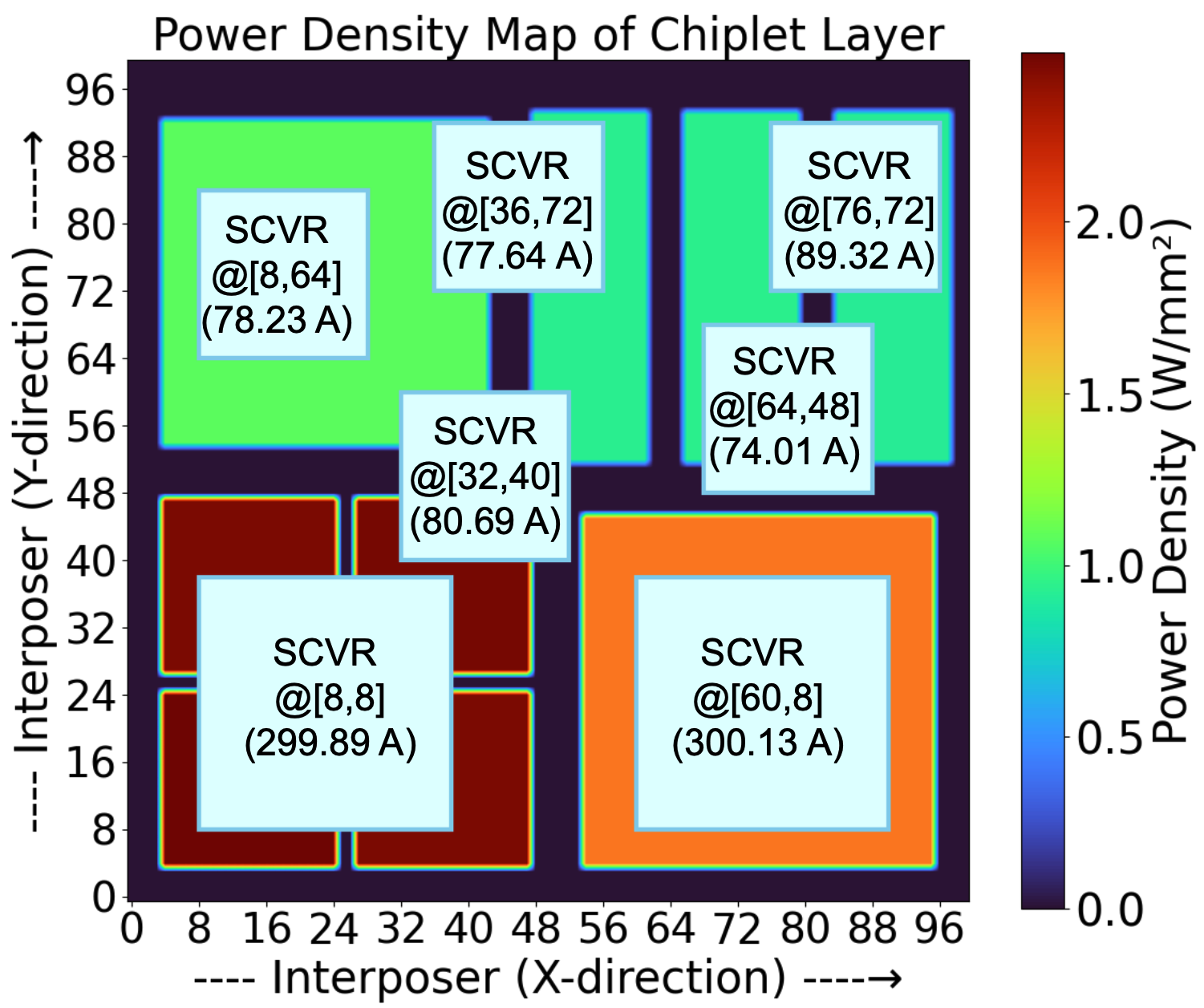}}
  \subfloat[]{\includegraphics[width=0.49\linewidth]{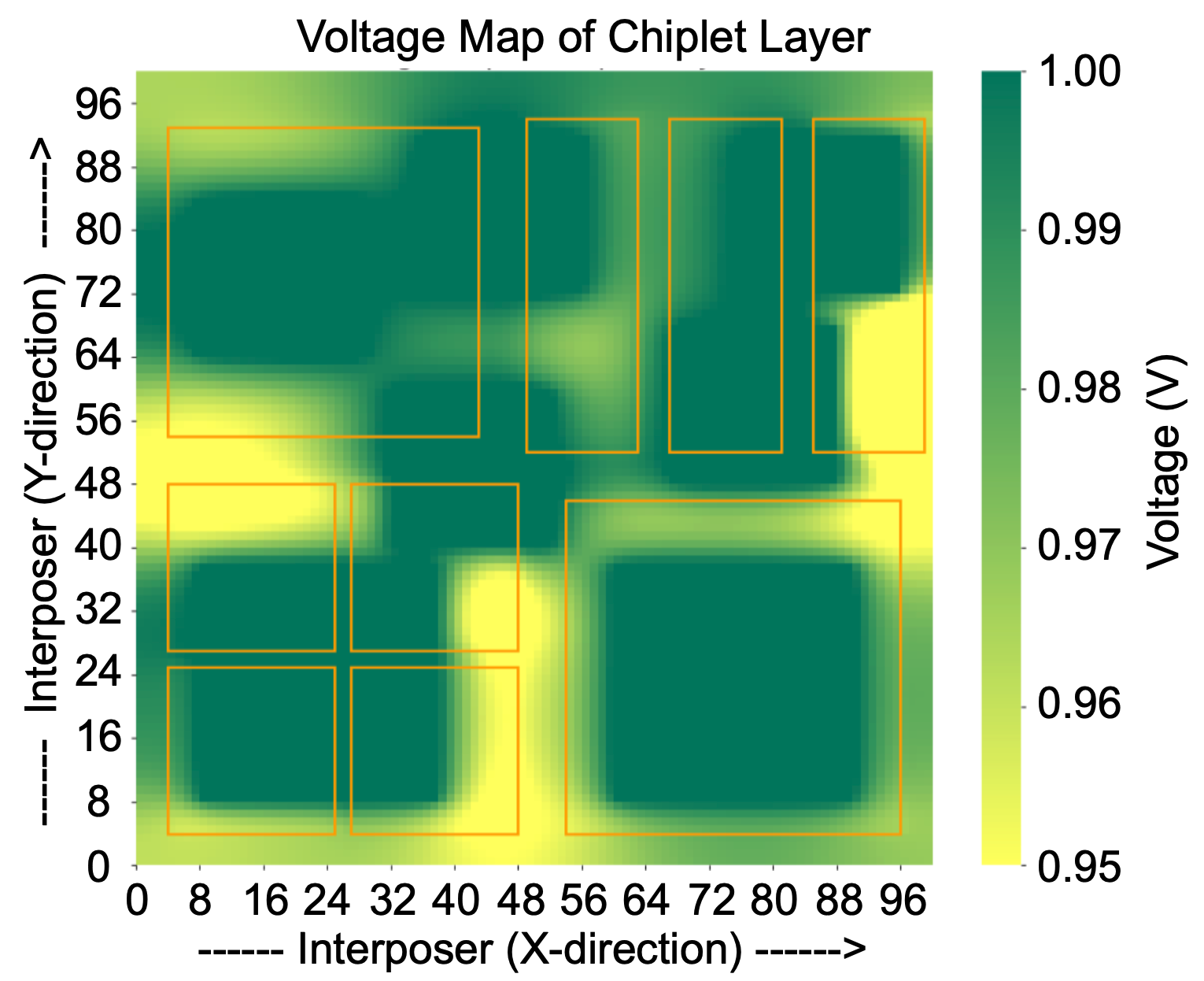}}

  \subfloat[]{\scriptsize
  \begin{tabular}{|c|c|c|c|c|}
    \hline
    \begin{tabular}[c]{@{}c@{}}Interposer    \\ Size (mm)             \\ \end{tabular} &
    \begin{tabular}[c]{@{}c@{}}Chiplet       \\ Load                  \\ (A) \end{tabular} &
    \begin{tabular}[c]{@{}c@{}}Power         \\ Density               \\ (W/mm$^2$) \end{tabular} &
    \begin{tabular}[c]{@{}c@{}}$I^t_{\text{SCVR,ideal}}$\\ 
                                               $t \in \{1,2\}$        \\ (A) \end{tabular} &
    \begin{tabular}[c]{@{}c@{}}Min. Voltage  \\ $V^{c}_{\text{spec}}$ \\ (V) \end{tabular} 
    \\ \hline
    \begin{tabular}[c]{@{}c@{}} 30$\times$30 \\ (Active area          \\ 634 mm$^2$) \end{tabular} & 
    \begin{tabular}[c]{@{}c@{}}50$\sim$300   \\ Total: 1000\end{tabular}  & 
    0.8$\sim$2.5 & 
    \begin{tabular}[c]{@{}c@{}}Type0: 310    \\ Type1: 100\end{tabular}  & 
    0.95 \\ \hline \multicolumn{5}{c}{}
  \end{tabular}
  }
  \vspace{-2mm}
  \caption{Voltage regulator placement for a testcase with heterogeneous chiplets: (a)~SCVR placement locations after optimization and power density map for the chiplet. (b)~Voltage map after placement. (c)~Chiplet/SCVR characteristics.}
  \label{fig:VRplacement}
  \vspace{-4mm}
\end{figure}

In the on-chip context, prior work has shown that the use of distributed structures, which reduce the distance from the regulator to the load relative to a single centralized regulator, are more effective in maintaining power levels at high efficiency~\cite{Gupta07,Zhou14,Wang17,VR_Place_2023}. In the HI context, there has been much less work in this area. The task of placing SCVRs must incorporate the objective of reducing losses in the PDN within the interposer. A large load on an SCVR draws a high output current, which produces a quadratic increase in DC power loss across the power delivery network~\cite{Vertical_PD_Inna}, leading to significant efficiency losses, as well as large voltage drops at the pins of the chiplets. However, placing too many SCVRs can result in large overheads. In~\cite{Zhang25}, based on macromodeling and a 0–1 mixed integer linear programming formulation, an approach is presented for finding the optimal number and locations of a set of SCVRs drawn from a library of multiple SCVR types.  This work primarily addresses DC loads and is appropriate for early-stage planning, but solutions for late-stage design and transient loads are as yet unexplored. An example solution from this approach is shown in Fig.~\ref{fig:VRplacement} for a heterogeneous testcase with chiplet current densities ranging from 0.8 to 2.5~A/mm$^2$. Two types of SCVRs are available, and the method finds the optimal position of the SCVRs that meets the voltage drop specification. The optimized locations of the SCVRs, and the corresponding voltage map, are shown in the figure.

Several open problems remain, e.g., (a)~regulator placement under transient loads; (b)~simultaneous optimization of the PDN topology, decap insertion, and regulator placement; (c)~regulator placement that is conscious of the regulation capabilities on individual chiplets; (d)~thermally aware placement that keeps regulators away from the hot spots; (e)~ML-assisted strategies for solving all of these problems.

\subsection{Decoupling capacitance insertion} 
\label{sec:decap}

\noindent
Several techniques are available to increase the reliability of power grids and control power grid noise, such as wire widening, grid topology optimization, and insertion of the decoupling capacitor (decap).  Of all these techniques, the decaps are arguably the most powerful method for reducing transient noise. Decaps serve as local charge reservoirs and can be used to satisfy sudden surges in current demand by the functional blocks/cells, while keeping supply voltage levels relatively stable.  Active/passive damping methods for resonant noise using decaps have also been proposed \cite{Gu07Jun,Xu07}.

Conventional decaps, implemented as SiO$_2$-based MOS capacitor structures, are widely used in robust power delivery network design.  3D power grid optimization has been studied in~\cite{Wong06,Huang07}. Unlike the 2D case, new considerations come into play while optimizing a 3D power grid using CMOS decaps, specifically related to congestion and leakage issues. The work in \cite{Zhou09} presents an approach for decap allocation in 3D power grids, using both conventional CMOS decaps and metal-insulator-metal (MIM) capacitors, which are fabricated between metal layers and have
high capacitance density and low leakage current density.  However, their use incurs the cost of routing blockages for signal or clock nets that attempt to cross them, and of the resistance of the via stack required to connect the current-drawing blocks in the device layer to the MIM capacitor. In \cite{Zhou09}, the decap budgeting problem, using both CMOS and MIM decaps, is formulated as a Linear Programming (LP) problem, and an efficient congestion-aware algorithm is proposed to optimize the power supply noise.   

HI substrates offer further options for embedded passives.  The use of deep trench capacitors provides even greater capacitance density, and arrays of such capacitors can be built into the HI substrate~\cite{Kannan20} and used as decaps. Decap placement in the HI substrate shares many characteristics of the on-chip decap placement problem, including the need for embedded decaps to be placed in proximity to high-powered chiplets, and the need to manage the routing blockages caused by decaps with the needs of clock and signal routes within the HI substrate.  Similarly, chiplet technologies that allow for on-chip embedded DRAM also use deep trench capacitors that are available for use as decaps.  A recent approach shows the application of reinforcement learning for decap placement in an HI substrate~\cite{Park20}.

Since a decap effectively supplies charge to feed a transient current demand, it is most effective when it is close to the current-drawing element. In~\cite{Popovich08, Todri09}, it is shown that a decap has a region of effectiveness: this is loosely referred to as a radius but the shape of the region is typically not circular and depends on the topology of the power grid. In advanced CMOS technology nodes, wires in lower metal layers have grown very narrow due to scaling, resulting in high per-unit-length resistances.   As a result, the region of effectiveness of an on-chiplet decap has shrunk, especially for fast current transients where the access path impedance limits charge delivery from distant decaps.  For decaps embedded in the HI substrate, the region of effectiveness is also bounded by the impedance through the substrate, chiplet interface, via stack, and on-die power grid. However, transitions in the HI substrate typically have lower frequency than those within a chiplet, and therefore they can be effective.

\subsection{Multistory power delivery in a 3D chiplet}
\label{sec:MSPD}

\begin{figure}[t]
\centering
\includegraphics[scale=0.3]{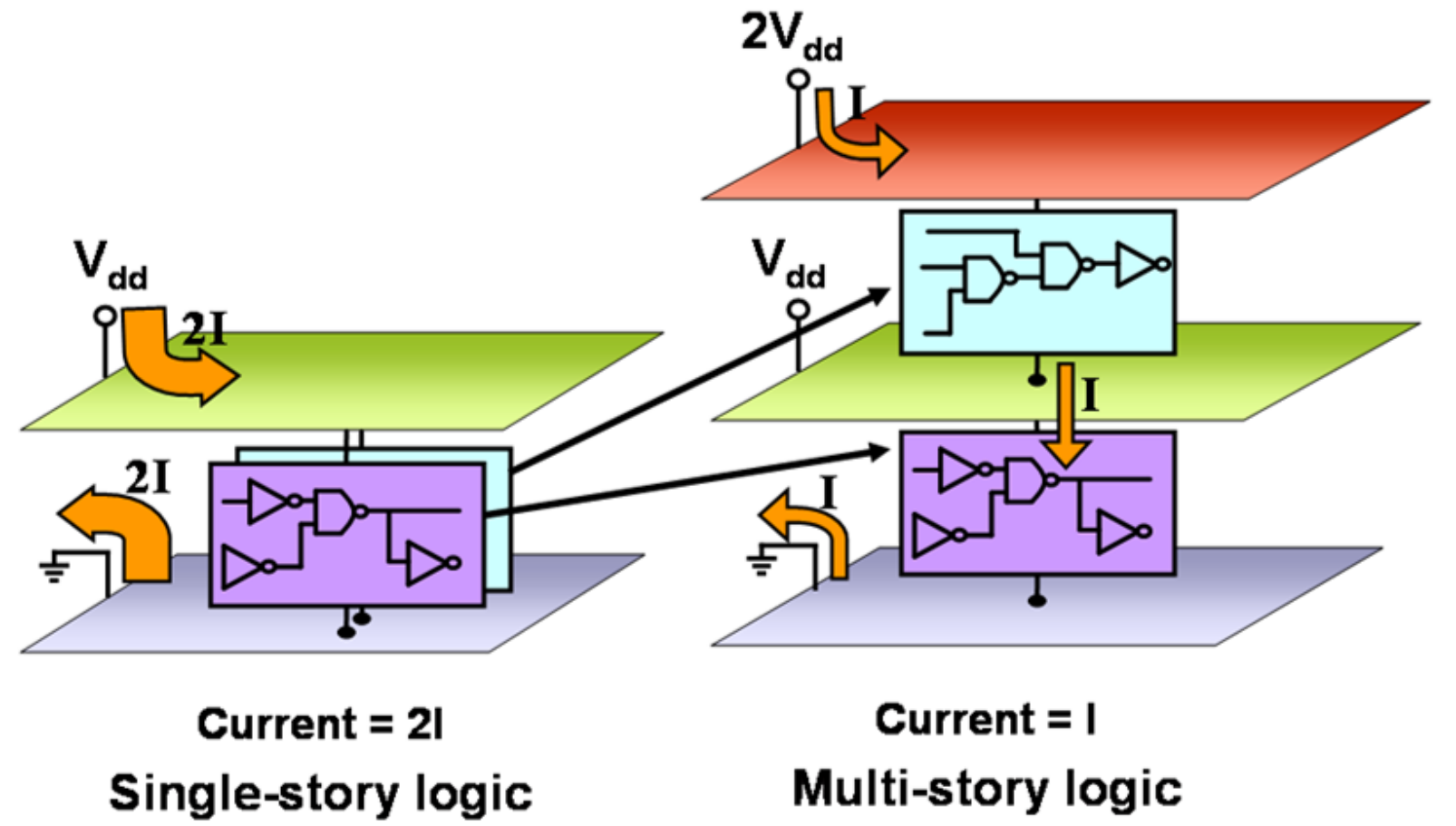}
\caption{Conventional and multistory power delivery schemes.}
\label{fig:PowDSche}
\vspace{-4mm}
\end{figure}

\noindent
In the context of 3D chiplets, the multistory power delivery (MSPD) scheme \cite{Gu05,Rajapandian05}, illustrated in Fig.~\ref{fig:PowDSche}, offers a potential route to efficient on-chip power delivery by stacking circuit blocks across a higher supply and recycling their current, which lowers the supply current, the power pin count, and the supply noise under a limited pin budget.  A schematic of a conventional supply network is shown in Fig.~\ref{fig:PowDSche}(a), where all circuits draw current from a single power source. Fig.~\ref{fig:PowDSche}(b) shows the multistory supply network, with subcircuits operating between two supply stories.  The concept of a ``story'' is merely an abstraction to illustrate the nature of the power delivery scheme, as opposed to the 3D IC architecture, where circuits are physically stacked in tiers.  In this scheme, current consumed in the 2V$_{dd}$--V$_{dd}$ story is subsequently recycled in the V$_{dd}$--Gnd story. Due to this internal recycling, half as much current is drawn compared to the conventional scheme, with almost the same total power consumption. A reduced current is beneficial since it cuts down the supply noise. Thus, in the best case, if the currents in the two subcircuits are completely balanced, the middle supply path will sink zero current. This results in minimal noise on that rail.  It has been demonstrated in \cite{Jain08} that the idea becomes particularly attractive for 3D IC structures involving stacked processors and memories.

The main issue with this technique is the requirement for separate body islands. This may be difficult in typical bulk processes. However, if we consider 3D ICs, the tiers are inherently separated electrically, which makes MSPD particularly attractive.  An important consideration in the design of an MSPD circuit is to locally maintain the current balance between logic blocks operating in different $V_{dd}$ domains. If balance is not maintained, the imbalance current cannot be recycled directly into the next story, and must instead either flow through the regulator on the middle supply path, or travel a long distance through significant parasitics to reach the next story, incurring significant power losses in either case. In an HI system with heterogeneous chiplets and dynamic AI workloads that step between near-zero and full load, maintaining this balance across stories is especially difficult, and any residual imbalance must be absorbed by the regulator on the middle supply path at the cost of efficiency.  Another important issue that has to be considered is the design stage at which the circuit should be partitioned into different $V_{dd}$ domains. Note that a level shifter is required at the output of a logic block if it is used to drive another logic block operating in a different $V_{dd}$ domain. Level-shifters occupy silicon area and cause extra delays in the circuit.

The MSPD idea can be automated by solving an optimization problem \cite{Zhan07}, which formulates the two-story problem as one of module assignment between the two stories.   The module assignment problem is addressed at the floorplanning level, where the module count is modest and block areas are abstracted out.  Experimental results demonstrate that the method is effective in building partitions for multistory power grids in both 2D and 3D structures, where blocks from multiple stories may coexist on the same tier or be on different tiers.  While not yet a mainstream approach, MSPD is a possible solution to the pin limitation problem in 3D HI systems.

However, in the 3D HI context, it is also important to recognize that the power delivery circuitry in the 3D stack is subject to high temperatures.  Disturbances in supply levels due to thermal effects at intermediate rails of one story can cause ripple effects in other stories of the stack. Sensitive blocks such as memory or analog circuitry are particularly susceptible to such disturbances, and these effects must therefore be accounted for when MSPD is contemplated.

\section{Outlook and Research Directions}
\label{sec:outlook}

\noindent
As stated in Section~\ref{sec:intro}, HI systems face tremendous challenges due to stringent requirements on delivering high power at high efficiency, across a large geometric scale but using microscopic components.  These systems require voltage regulation efficiencies that approach or exceed 90\%; the higher, the better. This requires holistic optimizations that allocate losses across all stages in Fig.~\ref{fig:architecture}.

A natural question is whether a high aggregate efficiency is achievable at all for these multi-kW systems, and which mechanisms ultimately place limits on achieving this goal. Today, Stage~1 can be built with efficiencies of 97\% or higher~\cite{Prakash24,Abramson21,Ahmed21,MPS23}:
assuming 98\% efficiency, this implies that the remainder of the system must operate at 91.8\% efficiency, so that the product of the two equals the desired 90\% goal. High overall efficiency involves making each component of the architecture as efficient as possible, and optimizing parameters that impact all components. Component-level losses fall into two categories:
\begin{itemize}
\item Losses in the \textit{distribution network}: These are resistive in nature, and when expressed as a fraction of the delivered power, they amount to the supply droop expressed as a fraction of the rail voltage. As long as the supply rail can delivery voltages to the load with low IR drops ($\sim$2\% of $V_{dd}$~\cite{Chhabria22-OpeNPDN}), the fractional distribution loss remains small.
\item Losses in the \textit{conversion stages}: These are harder to contain, and the limiting factor is in Stage~2 conversion to sub-1~V supply levels, where the conduction loss increases quadratically with the load current.  Moreover, the switching losses in the gate drivers, together with the bottom-plate and charge-sharing losses, impose a floor that becomes more severe as the conversion ratio grows.
\end{itemize}
An approach to loss allocation for one-stage structures, or for a fixed set of IBV values, has been proposed in~\cite{Krishnakumar26}, and represents an excellent starting point for tackling this issue.

Therefore, the critical bottlenecks lie in ensuring sufficiently low IR drop, which is a requirement for correct circuit operation, and in controlling losses in the Stage~2.  The first is likely to be achievable, leveraging intensive prior work on power delivery optimization at the chip and board levels. This is likely to require more efforts in investigating the use of distributed converters that not only meet IR drop requirements, but also ensure sufficiently fast response times to the bursty transients seen by modern workloads stemming from, for example, AI applications.  For the latter, it will be essential to use SCVRs in Stage~2.  Today's SCVR efficiencies approach close to 90\%~\cite{hardy_111_2023,Butzen23,Lu23,Assem22}, placing the goal of 90\% system efficiency within reach, and yet requiring the solution of several challenges.  The solution space encompasses silicon-based solutions, but is also enriched by allowing the use of non-silicon devices, which can be integrated into an HI substrate.

\begin{table*}[t]
\centering
\caption{List of open problems and critical research directions}
\label{tbl:challenges}
\begin{tabular}{|p{0.62in}|p{2in}|p{4in}|}
\cline{2-3}
\multicolumn{1}{c|}{} & \textbf{Problem} & \textbf{Suggested research directions} \\ \hline \hline
\multirow{4}{*}{\textbf{Near-term}}
& Small signal modeling of power converters, incorporating coupled inductor modeling and flying capacitor balancing 	(Section~\ref{sec:Stage1}) &
         Adapting averaged switch modeling (ASM) to determine transient performance of newly-proposed Stage 1 converters to optimize response time; incorporating models of coupled inductors into ASM; controlling flying capacitor voltages in hybrid converters as needed\\  \cline{2-3}
& Optimal IBV selection, incorporating load variations due to DVFS (Section~\ref{sec:IBV_selection})	&
        Modeling the impact of IBV on losses in Stage 1, Stage 2, and intermediate PDNs; using this model to optimize IBV subject to droop while maximizing system efficiency \\ \cline{2-3}
& PDN synthesis for the HI substrate,  matched to the on-chiplet PDNs	(Section~\ref{sec:PDN_opt}) &
        Adapting template-based PDN synthesis to HI-specific vertical stacks, RDLs,  bumps, vias, package pins, chiplet boundary models, and multiple voltage domains, incorporating resource contention between power, signal, and clock interconnects \\  \cline{2-3}
& Optimal regulator placement under transient and late-stage loads (Section~\ref{sec:VR_placement}) &
        Extending macromodel-based placement under static loads to the dynamic load case incorporating PDN impedance, regulator output impedance, control-loop response, decap insertation, and workload current traces \\  \hline \hline
\multirow{4}{*}{\textbf{Medium-term}}
& Power-delivery-aware chiplet selection during disaggregation (Section~\ref{sec:disaggregation}) &
        Incorporating power delivery capabilities of chiplets within a library to identify the optimal methods for disaggregating a system into a chiplet-based implementation \\ \cline{2-3}
& Decoupling capacitor insertion using embedded HI passives (Section~\ref{sec:decap}) &
        Adapting decap budgeting to embedded deep-trench and MIM capacitors while modeling region of effectiveness, equivalent series resistance/inductance, via resistance, routing blockage effects, leakage, and proximity to high slew load regions \\ \cline{2-3}
& Joint analysis of electrical performance, thermal impacts, and reliability consequences for power delivery components in an HI system (Sections~\ref{sec:thermal},~\ref{sec:reliability}) &
        Developing fast multiphysics-based methods for predicting temperature, performance, and electrical/mechanical reliability in HI systems in a coupled manner, using AI methods for early-stage prediction and fast physics-based analyses for late-stage analysis \\ \hline \hline
\multirow{4}{*}{\textbf{Long-term}}
& Power sharing optimization of PoL converters (Section~\ref{sec:Stage1})&
         Incorporating a lossless current sensing method for current equalization and/or droop control to prevent thermal runaway and ensure most efficient operation\\ \cline{2-3}
& Simultaneous optimization of PDN topology, decap insertion, and regulator placement (Sections~\ref{sec:PDN_opt}, \ref{sec:VR_placement}, \ref{sec:decap}) &
        Formulating a coupled multi-objective optimization over topology, regulator sites,  and decap allocation, using decomposition or surrogate-assisted search to handle electrical, thermal, area, and routing constraints \\ \cline{2-3}
& Modeling methodologies for worst-case  transient current workloads across the package-to-chip interface (Section~IV-C) &
        Developing workload-aware stimulus modeling that identifies realistic spatial and temporal worst-case excitations, capturing the bursty nature of challenging AI-related workloads for HI systems \\ \cline{2-3}
& Exploration of innovative methods for breaking through the bottleneck of pin count limitations for power delivery (Section~\ref{sec:MSPD}) &
        Exploring multi-story power delivery techniques with appropriate management of workload balance to limit losses associated with
        unrecyclable currents \\ \hline
\textbf{Continuing} & Open benchmarks and realistic synthetic testcases for HI &
        Driving the community to solve realistic problems through the release of open HI testcases with chiplet placement, power maps, voltage domains, package and substrate parasitics, regulator options, decap options, and validated electrical targets \\ \hline
\end{tabular}
\end{table*}

Table~\ref{tbl:challenges} summarizes the set of challenges and open problems required to meet the goals of HI power delivery, as articulated in this paper, together with a set of suggested research directions for overcoming them.  In the \textit{near term} (1--2 years), the first step is to move from individual block optimization to considering optimizations that holistically optimize the system.  Optimization of the IBV is a strong lever as it can link losses in various stages and can maximize overall efficiency.  Second, topology optimization within the HI substrate is a powerful knob for ensuring efficiency, and encompasses both PDN synthesis for the HI substrate and voltage regulator placement: early solutions in each area make these topics promising targets for near-term solutions.

In the \textit{medium term} (2--5 years), research that combines all elements of analysis to fully incorporate coupled performance, thermal, and reliability considerations will ensure system robustness in real environments.  Further, as new HI technologies are inducted into the mainstream, they will provide new methods for building decaps, notably through the use of DTCs. Research techniques must address methods for optimal decap placement, together with regulator placement, in the HI substrate, as well as the use of DTCs as flying capacitors within voltage regulators. In addition, as chiplet libraries become more mainstream, the problem of disaggregation to a library must consider chiplet characterization for power delivery as well as chiplet selection methods that incorporate on-chiplet voltage regulation as a design knob.

In the \textit{long term} (5--10 years), research must develop better methods for modeling worst-case power patterns, address the combination of techniques for increasingly integrated optimization of the power delivery system, incorporating static as well as transient analysis considerations.  It will also be necessary to explore unconventional methods for breaking through the barrier of limited pin counts, e.g., through the use of MSPD techniques, to meet the requirement to deliver high power within a limited footprint, and therefore a limited number of pins.  Finally, as a \textit{continuing} effort, good research requires good data for evaluation, and it is vital to seed efforts that provide open and realistic datasets on which researchers can report and evaluate their techniques.

\section{Conclusion}
\label{sec:conclusion}

\noindent
This paper describes the problems, challenges, and potential solution directions toward creating design methodologies for power delivery in 3D heterogeneously integrated systems.  The solution of problems in this area is fast-emerging as a major need for next-generation integrated systems, and today, there are no clear systematic methodologies or techniques that can be applied to general 3D HI systems. The intent of this article has been to list the problems and research opportunities, as well as, where available, summarizing related techniques that could be used to develop solutions to these problems in developing a power delivery architecture, in efficient analysis of these systems, and in optimization techniques that can reliably deliver power to these emerging multi-kW systems. This requires more than incremental adaptation.  Compared to 2D designs, the power densities are over an order of magnitude higher, and this is exacerbated by the pin limitation problem. While some solutions from 2D design can be adapted to specific subproblems, the problem requires rethinking of the entire power delivery solution, with greater amounts of distribution in the voltage regulation solutions, and taking into account thermal and reliability issues that arise in these systems.  In addition to developing new design techniques, an important issue to address is that of making testcases and benchmarks available to the research community, particularly for machine-learning-based solutions.  In this context, it will be important to generate realistic synthetic testcases, as has been done for PDNs in 2D design using machine learning techniques~\cite{Chhabria21BeGAN,Wu26}. Addressing all of these challenges will enable the community to develop rapid, reliable, and efficient design methodologies for 3D HI.

\section{Acknowledgments}

\noindent
The authors thank Divya Yogi and Yu (Kevin) Cao for helpful discussions on this topic.  

\bibliographystyle{misc/IEEEtran}
\bibliography{bib/main1,bib/main2,bib/References,bib/Ratullib-short}

@string{itpe = "IEEE Transactions on Power Electronics"}

@string{taes = "IEEE Transactions on Aerospace and Electronic Systems"}

@string{pesc = "IEEE Power Electronics Specialists Conference"}

@string{compelworkshop = "Proceedings of the IEEE Workshop on Control and Modeling for Power Electronics"}

@string{tie = "IEEE Transactions on Industrial Electronics"}

@string{tia = "IEEE Transactions on Industry Applications"}

@string{apec = "Proceedings of the IEEE Applied Power Electronics Conference and Exposition"}

@string{ecce = "Proceedings of the IEEE Energy Conversion Congress and Exposition"}

@string{intelec = "Proceedings of the International Telecommunications Energy Conference"}

@string{compel = "Proceedings of the Workshop on Control and Modeling for Power Electronics"}

@string{ojpe = "IEEE Open Journal of Power Electronics"}

@string{itia = "IEEE Transactions on Industry Applications"}

@string{itpe = "IEEE Trans. Power Electron."}

@string{taes = "IEEE Trans. Aerosp. Electron. Syst."}

@string{tia = "IEEE Trans. Ind. Appl."}

@string{pesc = "Proc. PESC"}

@string{compelworkshop = "Proc. COMPEL"}

@string{tie = "IEEE Trans. Ind. Electron."}

@string{apec = "Proc. APEC"}

@string{ecce = "Proc. ECCE"}

@string{intelec = "Proc. INTELEC"}

@string{compel = "Proc. COMPEL"}

@string{ojpe = "IEEE Open J. Power Electron."}

@string{itia = "IEEE Trans. Ind. Appl."}

@inproceedings{das_regulated_2019,
  title = {A Regulated {48V-to-1V}/100{A} 90.9\%-Efficient Hybrid Converter for {POL} Applications in Data Centers and Telecommunication Systems},
  booktitle = apec,
  author = {Das, Ratul and Le, Hanh-Phuc},
  year = 2019,
  pages = {1997--2001},
}

@article{wester_low-frequency_1973,
  title = {Low-Frequency Characterization of Switched dc-dc Converters},
  author = {Wester, G. W. and Middlebrook, R. D.},
  year = 1973,
  month = may,
  journal = taes,
  volume = {AES-9},
  number = {3},
  pages = {376--385},
}

@inproceedings{middlebrook_continuous_1975,
  title = {A Continuous Model for the Tapped-Inductor Boost Converter},
  booktitle = pesc,
  author = {Middlebrook, R. D.},
  year = 1975,
  pages = {63--79},
}

@inproceedings{urling_characterizing_1989,
  title = {Characterizing High-Frequency Effects in Transformer Windings -- {A} Guide to Several Significant Articles},
  booktitle = apec,
  author = {Urling, A. M. and Niemela, V. A. and Skutt, G. R. and Wilson, T. G.},
  year = 1989,
  pages = {373--385},
}

@techreport{cuk_modelling_1978,
  title = {Modelling, Analyses and Design of Switching Converters},
  author = {Cuk, S. M. and Middlebrook, R. D.},
  year = 1978,
  month = jan,
  institution = "NASA Lewis Research Center",
  address = "Cleveland, OH",
  number = {TRW-D04803-CFCM},
}

@article{shenoy_comparison_2016,
  title = {Comparison of a Buck Converter and a Series Capacitor Buck Converter for High-Frequency, High-Conversion-Ratio Voltage Regulators},
  author = {Shenoy, P. S. and Amaro, M. and Morroni, J. and Freeman, D.},
  year = 2016,
  month = oct,
  journal = itpe,
  volume = {31},
  number = {10},
  pages = {7006--7015},
}

@inproceedings{elasser_mini-lego_2023,
  title = {Mini-{LEGO}: A {1.5-MHz 240-A 48-V-to-1-V CPU VRM} with 8.4-mm height for vertical power delivery},
  booktitle = apec,
  author = {Elasser, Youssef and Baek, Jaeil and Radhakrishnan, Kaladhar and Gan, Houle and Douglas, Jonathan and De, Vivek and Jiang, Shuai and Krishnamurthy, Harish K. and Li, Xin and Sullivan, Charles R. and Chen, Minjie},
  year = 2023,
  pages = {1959--1966},
}

@inproceedings{das_averaged_2023,
  title = {Averaged Switch Modeling of Multi-Inductor Hybrid Converters},
  booktitle = compelworkshop,
  author = {Das, Ratul and Le, Hanh-Phuc},
  year = 2023,
  pages = {1--6},
}

@inproceedings{zhang_wide-bandwidth_2006,
  title = {Wide-Bandwidth Digital Multi-Phase Controller},
  booktitle = pesc,
  author = {Zhang, Yang and Zhang, Xu and Zane, Regan and Maksimovic, Dragan},
  year = 2006,
  note = "7 pages"
}

@inproceedings{kumar_high-performance_2018-1,
  title = {High-Performance Single-Stage Isolated {48V-to-1.8V} Point-of-Load Converter Utilizing Impedance Control Network and Distributed Transformer},
  booktitle = ecce,
  author = {Kumar, Ashish and Pervaiz, Saad and Afridi, Khurram K.},
  year = 2018,
  pages = {3838--3843},
}

@article{ahmed_single-stage_2020,
  title = {Single-Stage High-Efficiency 48/1~{V} Sigma Converter With Integrated Magnetics},
  author = {Ahmed, Mohamed H. and Fei, Chao and Lee, Fred C. and Li, Qiang},
  year = 2020,
  month = jan,
  journal = tie,
  volume = {67},
  number = {1},
  pages = {192--202},
}

@inproceedings{middlebrook_general_1976,
  title = {A General Unified Approach to Modelling Switching-Converter Power Stages},
  booktitle = pesc,
  author = {Middlebrook, R. D. and Cuk, Slobodan},
  year = 1976,
  pages = {18--34},
}

@article{vorperian_equivalent_1989,
  title = {Equivalent Circuit Models for Resonant and {PWM} Switches},
  author = {Vorperian, V. and Tymerski, R. and Lee, F.C.Y.},
  year = 1989,
  month = apr,
  journal = itpe,
  volume = {4},
  number = {2},
  pages = {205--214},
}

@article{vorperian_simplified_1990-1,
  title = {Simplified Analysis of {PWM} Converters Using Model of {PWM} Switch {Part~I: Continuous} Conduction Mode},
  author = {Vorperian, V.},
  year = 1990,
  month = may,
  journal = taes,
  volume = {26},
  number = {3},
  pages = {490--496},
}

@article{vorperian_simplified_1990-2,
  title = {Simplified Analysis of {PWM} Converters Using Model of {PWM} Switch {Part~II: Discontinuous} Conduction Mode},
  author = {Vorperian, V.},
  year = 1990,
  month = may,
  journal = taes,
  volume = {26},
  number = {3},
  pages = {497--505},
}

@article{middlebrook_null_1989,
  title = {Null Double Injection and the Extra Element Theorem},
  author = {Middlebrook, R. D.},
  year = 1989,
  month = aug,
  journal = {IEEE Trans. Educ.},
  volume = {32},
  number = {3},
  pages = {167--180},
}

@inproceedings{brown_sampled-data_1981,
  title = {Sampled-Data Modeling of Switching Regulators},
  booktitle = pesc,
  author = {Brown, Arthur R. and Middlebrook, R. D.},
  year = 1981,
  month = jun,
  pages = {349--369}
}

@inproceedings{lau_small-signal_1986,
  title = {Small-{{Signal Frequency Response Theory}} for Piecewise-Constant Two-Switched-Network Dc-to-Dc Converter Systems},
  booktitle = pesc,
  author = {Lau, Billy Y. and Middlebrook, R. D.},
  year = 1986,
  month = jun,
  pages = {186--200}
}

@misc{navitas_navitas_2026,
  title = {Navitas Debuts Revolutionary 800 {V}--6 {V} Power Delivery Board at {NVIDIA GTC} 2026},
  author = {{Navitas}},
  url = "https://navitassemi.com/",
  note = "{Accessed: April 27, 2026}",
  year = 2026,
  month = mar,
  urldate = {2026-04-27},
  langid = {american}
}

@misc{noauthor_epc_2025,
  title = {{{EPC}} 800 {VDC} power architecture for {AI} data centers},
  year = 2025,
  month = oct,
  journal = {Efficient Power Conversion Corporation},
  urldate = {2026-04-27},
  note = "{Accessed: April 27, 2026}",
  url = {https://epc-co.com/epc/about-epc/events-and-news/news/artmid/1627/articleid/3241/epc-800-vdc-power-architecture-for-ai-data-centers},
  langid = {american}
}

@misc{manners_ti_2026,
  title = {{{TI}} comes up with {{800V}} datacentre power architecture},
  author = {Manners, David},
  howpublished = "{\em Electronics Weekly}",
  year = 2026,
  month = mar,
  urldate = {2026-04-13},
  langid = {english}
}

@inproceedings{yeaman_datacenter_2007,
  title = {Datacenter power delivery architectures : {Efficiency} and annual operating costs},
  booktitle = {Darnell Digital Power Forum},
  author = {Yeaman, Paul},
  year = 2007,
  month = sep,
}

@inproceedings{cong_wang_quantitative_2014,
  title = {A Quantitative Comparison and Evaluation of {{48V DC}} and {{380V DC}} Distribution Systems for Datacenters},
  author = {Wang, Cong and Jain, Praveen},
  year = 2014,
  booktitle = intelec,
  pages = {1--7},
  urldate = {2018-07-02},
}

@inproceedings{fukui_hvdc_2010,
  title = {{{HVDC}} power distribution systems for telecom sites and data centers},
  booktitle = {Proc. ECCE-Asia},
  author = {Fukui, Akiyoshi and Takeda, Takashi and Hirose, Keiichi and Yamasaki, Mikio},
  year = 2010,
  pages = {874--880},
}

@article{ahmad_transition_2025,
  title = {The Transition from 54-{{V}} to 800-{{V}} Power in {{AI}} Data Centers},
  author = {Ahmad, Majeed},
  year = 2025,
  month = oct,
  journal = {EDN},
  note = "{Accessed: April 27, 2026}",
  url = "https://www.edn.com/the-transition-from-54-v-to-800-v-power-in-ai-data-centers/#google_vignette",
  langid = {american}
}

@misc{kumar_breaking_2026,
  title = {Breaking the 12-{{V}} Bottleneck: The Shift to 48-{{V}} Power},
  author = {Kumar, Rakesh},
  year = 2026,
  month = nov,
  urldate = {2026-04-27},
  howpublished = "{\em Power Electronic Tips}",
  url = "https://www.powerelectronictips.com/breaking-the-12-v-bottleneck-the-shift-to-48-v-power/",
  note = "{Accessed: April 27, 2026}"
}

@inproceedings{lyu_composite_2018,
  title = {Composite modular power delivery architecture for next-gen {48V} data center applications},
  booktitle = "Proc. WiPDA Asia",
  author = {Lyu, Xiaofeng and Li, Yanchao and Ni, Ze and Johnson, Jalen and Cao, Dong and Nan, Chenhao and Jiang, Shuai},
  year = 2018,
  pages = {343--350},
  doi = {10.1109/WiPDAAsia.2018.8734572},
}

@misc{li_google_2019,
  title = {Google {48V} rack adaptation and onboard power technology update},
  author = {Li, Xin and Jiang, Shuai},
howpublished = "{\em Open Compute Project Summit}",
  year = 2019,
  month = mar,
  langid = {english}
}

@inproceedings{baek_lego-pol_2020,
  title = {{{LEGO-PoL}}: {{A 48V-1.5V 300A} merged-two-stage hybrid converter for ultra-high-current microprocessors}},
  booktitle = apec,
  author = {Baek, Jaeil and Wang, Ping and Elasser, Youssef and Chen, Yenan and Jiang, Shuai and Chen, Minjie},
  year = 2020,
  pages = {490--497},
}

@inproceedings{baek_lego-pol_2019,
  title = {{LEGO-PoL}: {A} 93.1\% {54V-1.5V 300A} merged-two-stage hybrid converter with a linear extendable group operated point-of-load ({{LEGO-PoL}}) architecture},
  booktitle = compel,
  author = {Baek, Jaeil and Wang, Ping and Jiang, Shuai and Chen, Minjie},
  year = 2019,
  note = {8 pages},
}

@article{jiang_switched_2019,
  title = {Switched tank converters},
  author = {Jiang, Shuai and Saggini, Stefano and Nan, Chenhao and Li, Xin and Chung, Chee and Yazdani, Mobashar},
  year = 2019,
  journal = itpe,
  volume = {34},
  number = {6},
  pages = {5048--5062},
  issn = {0885-8993, 1941-0107},
  doi = {10.1109/TPEL.2018.2868447}
}

@article{li_98.55_2018,
  title = {A 98.55\% efficiency switched-tank converter for data center application},
  author = {Li, Y. and Lyu, X. and Cao, D. and Jiang, S. and Nan, C.},
  year = 2018,
  journal = tia,
  pages = {1--1},
  issn = {0093-9994},
  doi = {10.1109/TIA.2018.2858741}
}

@inproceedings{zhu_family_2019,
  title = {A family of transformerless stacked active bridge converters},
  booktitle = apec,
  author = {Zhu, Jianglin and Maksimovi{\'c}, Dragan},
  year = 2019,
  pages = {19--24},
}

@inproceedings{zhu_dickson-squared_2022,
  title = {A {Dickson}-Squared Hybrid Switched-Capacitor Converter for Direct 48 {{V}} to Point-of-Load Conversion},
  booktitle = apec,
  author = {Zhu, Yicheng and Ge, Ting and Ye, Zichao and {Pilawa-Podgurski}, Robert C.N.},
  year = 2022,
  month = mar,
  pages = {1272--1278},
  issn = {2470-6647},
  doi = {10.1109/APEC43599.2022.9773567}
}

@article{zhu_comparative_2024,
  title = {Comparative Performance Analysis of Regulated Hybrid Switched-Capacitor Topologies for Direct 48 {{V}} to Point-of-Load Conversion},
  author = {Zhu, Yicheng and Ellis, Nathan MILES and {Pilawa-Podgurski}, Robert C. N.},
  year = 2024,
  journal = ojpe,
  volume = {5},
  pages = {1735--1755},
  issn = {2644-1314},
  doi = {10.1109/OJPEL.2024.3478288},
  urldate = {2025-04-17}
}

@misc{baba_benefits_2012,
  title = {Benefits of a Multiphase Buck Converter},
  author = {Baba, David},
  year = 2012,
  url = "https://www.ti.com/lit/an/slyt449/slyt449.pdf?ts=1777287451620",
  note = "8 pages, {Accessed: April 27, 2026}",
  institution = {Texas Instruments Inc.},
  langid = {english}
}

@techreport{parisi_multiphase_2017,
  title = {Multiphase buck design from start to finish ({Part} 1)},
  author = {Parisi, Carmen},
  year = 2017,
  url = "https://www.ti.com/lit/an/slva882b/slva882b.pdf",
  note = "18 pages, {Accessed: April 27, 2026}",
  institution = {Texas Instruments Inc.},
  langid = {english}
}

@phdthesis{das_topologies_2022,
  title = {Topologies, modeling, and control of hybrid switched-capacitor converters},
  author = {Das, Ratul},
  year = 2022,
  month = dec,
  langid = {english},
  school = {University of California San Diego},
  address = "San Diego, CA"
}

@phdthesis{kumar_high-performance_2018,
  title = {High-performance power converters for telecom and datacenter applications},
  author = {Kumar, Ashish},
  year = 2018,
  school = {University of Colorado Boulder},
  address = "Boulder, CO"
}

@inproceedings{ahmed_high-efficiency_2017,
  title = {High-Efficiency High-Power-Density 48/{{1V}} Sigma Converter Voltage Regulator Module},
  booktitle = apec,
  author = {Ahmed, M. and Fei, C. and Lee, F. C. and Li, Q.},
  year = 2017,
  pages = {2207--2212},
  address = {Tampa, FL},
  doi = {10.1109/APEC.2017.7931005}
}

@inproceedings{ahmed_startup_2017,
  title = {Startup and control of high efficiency 48/{{1V}} sigma converter},
  booktitle = ecce,
  author = {Ahmed, M. H. and Fei, C. and Li, V. and Lee, F. C. and Li, Q.},
  year = 2017,
  pages = {2010--2016},
}

@inproceedings{hardy_111_2023,
  title = {A Scalable Heterogeneous Integrated Two-Stage Vertical Power-Delivery Architecture for High-Performance Computing},
  booktitle = isscc,
  author = {Hardy, Casey and Pham, Hieu and Jatlaoui, Mohamed Mehdi and Voiron, Frederic and Xie, Tianshi and Chen, Po-Han and Jha, Saket and Mercier, Patrick and Le, Hanh-Phuc},
  year = 2023,
  pages = {182--184},
}

@inproceedings{nabih_low-profile_2021,
  title = {Low-Profile and High-Efficiency 3 {kW} 400 {V-48 V LLC} Converter with a Matrix of Four Transformers and Inductors for {48V} Power Architecture for Data Centers},
  booktitle = ecce,
  author = {Nabih, Ahmed and Li, Qiang},
  year = 2021,
  pages = {1813--1819},
}

@inproceedings{fu_10mhz_2026,
  title = {{10MHz 48V-12V DCX} With {6100W}/in$^3$ Power Density for Future Data Center Applications},
  booktitle = apec,
  author = {Fu, Chaowei and Wu, Xinke and Zhang, Junming},
  year = 2026,
  pages = {126--131},
}

@inproceedings{winkler_increasing_2026,
  title = {Increasing Power Density in {48V DC-DC} Converters Using Vertically Integrated {3D} Glass Power Bricks for Data Center Applications},
  booktitle = apec,
  author = {Winkler, Joseph and Rumpf, Torben and Koch, Jannik and Wurz, Marc Christopher and Wicht, Bernhard},
  year = 2026,
  pages = {1072--1076},
}

@inproceedings{fang_961_2025,
  title = {A 96.1\% Efficiency {48V-to-IBV GaN} Power Converter with Full-Wave Temperature-Compensated Current Sensing and Adaptive Slope Emulation Achieving 4.3\% Full-Temperature Sensing Error for {AI} Data Center Applications},
  booktitle = cicc,
  author = {Fang, Yike and Zou, Jie and Ke, Xugang and He, Lenian},
  year = 2025,
  note = "3 pages",
}

@article{elasser_mini-lego_2024,
  title = {Mini-{LEGO CPU} Voltage Regulator},
  author = {Elasser, Youssef and Baek, Jaeil and Radhakrishnan, Kaladhar and Gan, Houle and Douglas, Jonathan P. and Krishnamurthy, Harish K. and Li, Xin and Jiang, Shuai and De, Vivek and Sullivan, Charles R. and Chen, Minjie},
  year = 2024,
  month = mar,
  journal = itpe,
  volume = {39},
  number = {3},
  pages = {3391--3410},
}

@article{zhu_switching_2026,
  title = {A Switching Bus Converter Enabling Direct 48-{V}-to-Point-of-Load Vertical Power Delivery for High-Performance Data Center Processors},
  author = {Zhu, Yicheng and Ellis, Nathan M. and Kudva, Sudhir S. and Mosa, Mostafa and Gray, C. Thomas and {Pilawa-Podgurski}, Robert C. N.},
  year = 2026,
  month = jul,
  journal = itpe,
  volume = {41},
  number = {7},
  pages = {11098--11119},
}

@inproceedings{zhu_500-48--1-v_2023,
  title = {A 500-{A}/48-to-1-{V} Switching Bus Converter: A Hybrid Switched-Capacitor Voltage Regulator with 94.7\% Peak Efficiency and 464-{W}/in$^3$ Power Density},
  shorttitle = {A 500-{A}/48-to-1-{V} Switching Bus Converter},
  booktitle = apec,
  author = {Zhu, Yicheng and Ge, Ting and Ellis, Nathan M. and Horowitz, Logan and {Pilawa-Podgurski}, Robert C. N.},
  year = 2023,
  pages = {1989--1996},
}

@article{wu_hybrid_2024,
  title = {Hybrid Resonant Converter-Based 8:1 Bus Converter With 3.5 {kW/in$^3$} and 98.6\%-Efficient for 48 {V} Data-Center Power Systems},
  shorttitle = {Hybrid {{Resonant Converter-Based}} 8},
  author = {Wu, Hongfei and Zhang, Yu and Li, Zewei},
  year = 2024,
  month = jan,
  journal = itpe,
  volume = {39},
  number = {1},
  pages = {36--41},
}

@article{figueroa_2200a48v--1v_2026,
  title = {2200{A}/{48V-to-1V} Low-Profile Direct Power Converter with Standard {PCB} Transformer},
  author = {Figueroa, Alejandro and Mazariegos, Pablo and Cobos, {\'A}lvaro and Goicoechea, Javier and Castro, Alejandro and Delgado, Alberto and Cobos, Jos{\'e} A.},
  year = 2026,
  journal = itpe,
  note = "(early access), 14 pages"
}

@inproceedings{rentmeister_48v:2v_2017,
  title = {A {{48V}}:{{2V}} Flying Capacitor Multilevel Converter Using Current-Limit Control for Flying Capacitor Balance},
  booktitle = apec,
  author = {Rentmeister, J. S. and Stauth, J. T.},
  year = 2017,
  pages = {367--372},
}

@article{stillwell_active_2019,
  title = {Active Voltage Balancing in Flying Capacitor Multi-Level Converters With Valley Current Detection and Constant Effective Duty Cycle Control},
  author = {Stillwell, A. and Candan, E. and {Pilawa-Podgurski}, R. C. N.},
  year = 2019,
  month = nov,
  journal = itpe,
  volume = {34},
  number = {11},
  pages = {11429--11441},
  issn = {1941-0107},
}

@inproceedings{xia_state_2019,
  title = {State Space Analysis of Flying Capacitor Multilevel {DC-DC} Converters for Capacitor Voltage Estimation},
  booktitle = apec,
  author = {Xia, Ziyu and Dobbins, Benjamin and Rentmeister, J. S. and Stauth, J. T.},
  year = 2019,
}

@inproceedings{celikovic_modeling_2019,
  title = {Modeling of Capacitor Voltage Imbalance in Flying Capacitor Multilevel {DC-DC} Converters},
  booktitle = compel,
  author = {Celikovic, Janko and Das, Ratul and Le, Hanh-Phuc and Maksimovic, Dragan},
  year = 2019,
  note = "8 pages"
}

@inproceedings{das_demystifying_2019,
  title = {Demystifying Capacitor Voltages and Inductor Currents in Hybrid Converters},
  booktitle = compel,
  author = {Das, Ratul and Celikovic, Janko and Abedinpour, Siamak and Mercer, Mark and Maksimovic, Dragan and Le, Hanh-Phuc},
  year = 2019,
  note = "8 pages",
}

@article{zhou_balancing_2024,
  title = {Balancing Multiphase {FCML} Converters With Coupled Inductors: Modeling, Analysis, Limitations},
  author = {Zhou, Daniel H. and {\v C}elikovi{\'c}, Janko and Maksimovi{\'c}, Dragan and Chen, Minjie},
  year = 2024,
  month = aug,
  journal = itpe,
  volume = {39},
  number = {8},
  pages = {9268--9291},
}

@article{wang_matrix_2022,
  title = {Matrix {{Coupled All-in-One Magnetics}} for {{PWM Power Conversion}}},
  author = {Wang, Ping and Zhou, Daniel H. and Elasser, Youssef and Baek, Jaeil and Chen, Minjie},
  year = 2022,
  month = dec,
  journal = itpe,
  volume = {37},
  number = {12},
  pages = {15035--15050},
}

@article{qi_comparative_2024,
  title = {Comparative Study of Four Droop Control Strategies in Buck Converter Based {DC} Microgrid},
  author = {Qi, Li Lisa and Gao, Min and Faddel, Samy},
  year = 2024,
  month = jul,
  journal = itia,
  volume = {60},
  number = {4},
}

@book{rincon-mora_switched_2023,
  title = {Switched Inductor Power IC Design},
  author = {{Rinc{\'o}n-Mora}, Gabriel Alfonso},
  year = 2023,
  publisher = {Springer},
  address = {Cham, Switzerland},
}

@string{cad = "IEEE Transactions on Computer-Aided Design of Integrated Circuits and Systems"}

@string{cas = "IEEE Transactions on Circuits and Systems"}

@string{cas1 = "IEEE Transactions on Circuits and Systems I"}

@string{tvlsi = "IEEE Transactions on VLSI Systems"}

@string{cpmt = "IEEE Transactions on Components and Packaging Technologies"}

@string{tdmr = "IEEE Transactions on Devices and Materials Reliability"}

@string{jssc = "IEEE Journal of Solid-State Circuits"}

@string{iccad = "Proceedings of the IEEE/ACM International Conference on Computer-Aided Design"}

@string{dac = "Proceedings of the ACM/IEEE Design Automation Conference"}

@string{aspdac = "Proceedings of the Asia-South Pacific Design Automation Conference"}

@string{date = "Proceedings of the Design, Automation \& Test in Europe"}

@string{cicc = "Proceedings of the IEEE Custom Integrated Circuits Conference"}

@string{isscc = "Proceedings of the IEEE International Solid-State Circuits Conference"}

@string{isqed = "Proceedings of the IEEE International Symposium on Quality Electronic Design"}

@string{ispd = "Proceedings of the International Symposium on Physical Design"}

@string{islped = "Proceedings of the ACM International Symposium on Low Power Electronics and Design"}

@string{iedm = "Proceedings of the IEEE International Electron Devices Meeting"}

@string{todaes = "ACM Transactions on Design Automation of Electronic Systems"}

@string{ectc = "Proceedings of the IEEE International Conference on Electronics Components and Technology"}

@string{mr="Microelectronics Reliability"}

@string{me="Microelectronics Engineering"}

@string{bcicts = "Proceedings of the IEEE BiCMOS and Compound Semiconductor Integrated Circuits and Technology Symposium"}

@string{svc = "Proceedings of the IEEE Symposium on VLSI Circuits"}

@string{tcad = "IEEE Transactions on Computer-Aided Design of Integrated Circuits and Systems"}

@string{hpca = "Proceedings of the IEEE International Symposium on High-Performance Computer Architecture"}

@string{asscc = "Proceedings of the IEEE Asian Solid-State Circuits Conference"}

@string{jetcas = "IEEE Journal on Emerging and Selected Topics in Circuits and Systems"}

@string{mlcad = "Proceedings of the ACM/IEEE International Symposium on Machine Learning for CAD"}

@string{epep = "Proceedings of Electrical Performance of Electronic Packaging"}

@string{svtc = "Proceedings of the IEEE Symposium on VLSI Technology and Circuits"}

@string{cad = "IEEE Trans. Comput. Aided Des. Integr. Circuits Syst."}

@string{tdmr = "IEEE Trans. Device Mater. Reliab."}

@string{jssc = "IEEE J. Solid-State Circuits"}

@string{iccad = "Proc. ICCAD"}

@string{dac = "Proc. DAC"}

@string{aspdac = "Proc. ASP-DAC"}

@string{date = "Proc. DATE "}

@string{cicc = "Proc. CICC"}

@string{mr = "Microelectron. Reliab."}

@string{me="Microelectron. Eng."}

@string{isqed = "Proc. ISQED"}

@string{ispd = "Proc. ISPD"}

@string{iedm = "Proc. IEDM"}

@string{todaes = "ACM Trans. Des. Autom. Electron. Syst."}

@string{ectc = "Proc. ECTC"}

@string{tcad = "IEEE Trans. Comput. Aided Des. Integr. Circuits Syst."}

@string{tvlsi = "IEEE Trans. VLSI Syst."}

@string{cas1 = "IEEE Trans. Circuits Syst. I Regul. Pap."}

@string{micro="Proc. MICRO"}

@string{isscc = "Proc. ISSCC"}

@string{mlcad = "Proc. MLCAD"}

@string{hotchips = "Proc. Hot Chips"}

@string{apr = "Appl. Mech. Rev."}

@string{apr = "Appl. Phys. Rev."}

@string{mr = "Microelectron. Rel."}

@string{me = "Microelectron. Eng."}

@string{jvstb = "J. Vac. Sci. Technol., B"}

@string{cpmt = "IEEE Trans. Compon. Packag. Technol."}

@string{ijhmt = "Int. J. Heat Mass Transfer"}

@string{bcicts = "Proc. BCICTS"}

@string{amr = "Appl. Mech. Rev."}

@string{socc = "Proc. SOCC"}

@string{fnt = "Found. Trends in Electron. Des. Autom."}

@string{asscc = "Proc. ASSCC"}

@string{jetcas = "IEEE J. Emerging Sel. Top. Circuits Syst."}

@string{hpca = "Proc. HPCA"}

@string{epep = "Proc. EPEP"}

@string{svc = "Proc. Symp. VLSI Circuits"}

@string{svtc = "Proc. Symp. VLSI Technol. Circuits"}

@string{islped = "Proc. ISLPED"}

@ARTICLE{Converter_sizing_Julien,
  author={De Vos, Julien and Flandre, Denis and Bol, David},
  journal= cas1, 
  title="A Sizing Methodology for On-Chip Switched-Capacitor {DC-DC} Converters", 
  year={2014},
  volume={61},
  number={5},
  pages={1597-1606},
}

@INPROCEEDINGS{Verti_PDinHI_HanhPhuc,
  author={Le, Hanh-Phuc and Hardy, Casey and Pham, Hieu and Jatlaoui, Mohamed Mehdi and Voiron, Frederic and Mercier, Patrick and Chen, Po-Han and Jha, Saket},
  booktitle= bcicts, 
  title="Vertical Power Delivery and Heterogeneous Integration for High-Performance Computing", 
  year={2023},
  pages={32-35},
  doi={10.1109/BCICTS54660.2023.10310708}}

@INPROCEEDINGS{Integrated_PDin3D_BorisVaisband,
  author={Safari, Yousef and Vaisband, Boris},
  booktitle= isqed, 
  title="Integrated Power Delivery Methodology for {3D ICs}", 
  year={2022},
  volume={},
  number={},
  pages={114-119},
  doi={10.1109/ISQED54688.2022.9806286}}

@ARTICLE{PD_for_HPMicropro_Kaladhar,
  author={Radhakrishnan, Kaladhar and Swaminathan, Madhavan and Bhattacharyya, Bidyut K.},
  journal=cpmt, 
  title="Power Delivery for High-Performance Microprocessors -- {C}hallenges, Solutions, and Future Trends", 
  year={2021},
  volume={11},
  number={4},
  pages={655-671},
  doi={10.1109/TCPMT.2021.3065690}}

@ARTICLE{INTACT_Vivet_Pascal,
  author={Vivet, Pascal and Guthmuller, Eric and Thonnart, Yvain and Pillonnet, Gael and Fuguet, César and Miro-Panades, Ivan and Moritz, Guillaume and Durupt, Jean and Bernard, Christian and Varreau, Didier and Pontes, Julian and Thuries, Sébastien and Coriat, David and Harrand, Michel and Dutoit, Denis and Lattard, Didier and Arnaud, Lucile and Charbonnier, Jean and Coudrain, Perceval and Garnier, Arnaud and Berger, Frédéric and Gueugnot, Alain and Greiner, Alain and Meunier, Quentin L. and Farcy, Alexis and Arriordaz, Alexandre and Chéramy, Séverine and Clermidy, Fabien},
  journal=jssc, 
  title="{IntAct}: A 96-Core Processor with Six Chiplets 3{D}-Stacked on an Active Interposer with Distributed Interconnects and Integrated Power Management", 
  year={2021},
  volume={56},
  number={1},
  pages={79-97},
  doi={10.1109/JSSC.2020.3036341}}

@INPROCEEDINGS{Lakefield_3D,
  author={Gomes, Wilfred and Khushu, Sanjeev and Ingerly, Doug B. and Stover, Patrick N. and Chowdhury, Nasirul I. and O'Mahony, Frank and Balankutty, Ajay and Dolev, Noam and Dixon, Martin G. and Jiang, Lei and Prekke, Surya and Patra, Biswajit and Rott, Pavel V. and Kumar, Rajesh},
  booktitle=isscc, 
  title="Lakefield and Mobility Compute: A 3{D} Stacked 10nm and 22{FFL} Hybrid Processor System in 12×12mm$^2$, 1mm Package-on-Package", 
  year={2020},
  volume={},
  number={},
  pages={144-146},
  doi={10.1109/ISSCC19947.2020.9062957}}

@ARTICLE{CoWoS_3D,
  author={Lin, Mu-Shan and Huang, Tze-Chiang and Tsai, Chien-Chun and Tam, King-Ho and Hsieh, Cheng-Hsiang and Chen, Tom and Huang, Wen-Hung and Hu, Jack and Chen, Yu-Chi and Goel, Sandeep Kumar and Fu, Chin-Ming and Rusu, Stefan and Li, Chao-Chieh and Yang, Sheng-Yao and Wong, Mei and Yang, Shu-Chun and Lee, Frank},
  journal=jssc, 
  title="A 7nm 4{GHz} {Arm}-Core-Based {CoWoS} Chiplet Design for High-Performance Computing", 
  year={2020},
  volume={55},
  number={4},
  pages={956-966},
  doi={10.1109/JSSC.2019.2960207}}

@book{Matrix_Comp_Golub_2013,
  title="Matrix Computations",
  author={Golub, G. H. and Van Loan, C. F.},
  isbn={9781421407944},
  year={2013},
  publisher={Johns Hopkins University Press},
  address={Baltimore, MD}
}

@INPROCEEDINGS{Intel_EMIB,
  author={Mahajan, Ravi and Sankman, Robert and Patel, Neha and Kim, Dae-Woo and Aygun, Kemal and Qian, Zhiguo and Mekonnen, Yidnekachew and Salama, Islam and Sharan, Sujit and Iyengar, Deepti and Mallik, Debendra},
  booktitle=ectc, 
  title="Embedded Multi-die Interconnect Bridge ({EMIB}) -- A High Density, High Bandwidth Packaging Interconnect", 
  year={2016},
  volume={},
  number={},
  pages={557-565},
  doi={10.1109/ECTC.2016.201}}

@INPROCEEDINGS{Intel_Foveros,
  author={Ingerly, D. B. and Amin, S. and Aryasomayajula, L. and Balankutty, A. and Borst, D. and Chandra, A. and Cheemalapati, K. and Cook, C. S. and Criss, R. and Enamul, K. and Gomes, W. and Jones, D. and Kolluru, K. C. and Kandas, A. and Kim, G.-S. and Ma, H. and Pantuso, D. and Petersburg, C.F. and Phen-givoni, M. and Pillai, A. M. and Sairam, A. and Shekhar, P. and Sinha, P. and Stover, P. and Telang, A. and Zell, Z.},
  booktitle=iedm, 
  title="Foveros: {3D} Integration and the Use of Face-to-Face Chip Stacking for Logic Devices", 
  year={2019},
  volume={},
  number={},
  pages={19.6.1-19.6.4},
  doi={10.1109/IEDM19573.2019.8993637}}

@INPROCEEDINGS{Edward_FIVR,
  author={Burton, Edward A. and Schrom, Gerhard and Paillet, Fabrice and Douglas, Jonathan and Lambert, William J. and Radhakrishnan, Kaladhar and Hill, Michael J.},
  booktitle=apec, 
  title="{FIVR} -- Fully Integrated Voltage Regulators on 4th Generation {Intel® Core™ SoCs}", 
  year={2014},
  volume={},
  number={},
  pages={432-439},
  doi={10.1109/APEC.2014.6803344}}

@ARTICLE{Anderson_pareto_opt_of_SCVRs,
  author={Andersen, Toke M. and Krismer, Florian and Kolar, Johann W. and Toifl, Thomas and Menolfi, Christian and Kull, Lukas and Morf, Thomas and Kossel, Marcel and Brändli, Matthias and Francese, Pier Andrea},
  journal=itpe, 
  title="Modeling and {Pareto} Optimization of On-Chip Switched Capacitor Converters", 
  year={2017},
  volume={32},
  number={1},
  pages={363-377},
  doi={10.1109/TPEL.2016.2529501}}

@ARTICLE{Hanh_DesignTech_of_SCVR,
  author={Le, Hanh-Phuc and Sanders, Seth R. and Alon, Elad},
  journal=jssc, 
  title="Design Techniques for Fully Integrated Switched-Capacitor {DC-DC} Converters", 
  year={2011},
  volume={46},
  number={9},
  pages={2120-2131},
  doi={10.1109/JSSC.2011.2159054}}

@ARTICLE{Seeman_Analysis_optimze_SCVRs,
  author={Seeman, Michael D. and Sanders, Seth R.},
  journal=itpe, 
  title="Analysis and Optimization of Switched-Capacitor {DC-DC} Converters", 
  year={2008},
  volume={23},
  number={2},
  pages={841-851},
}

@INPROCEEDINGS{Sudarshan24,
  author={Sudarshan, Chetan Choppali and Matkar, Nikhil and Vrudhula, Sarma and Sapatnekar, Sachin S. and Chhabria, Vidya A.},
  booktitle=hpca,
  title={{ECO-CHIP}: Estimation of Carbon Footprint of Chiplet-based Architectures for Sustainable {VLSI}}, 
  year={2024},
  pages={671-685},
}

@inproceedings{Nalla25,
author = "P. S. Nalla and E. Haque and Y. Liu and S. S. Sapatnekar and J. Zhang and C.
Chakrabarti and Y. Cao",
title = "{CLAIRE}: Composable Chiplet Libraries for {AI} Inference",
booktitle = date,
year = 2025,
note="6 pages"
}

@ARTICLE{Chhabria22-OpeNPDN,
  author={Chhabria, V. A. and Sapatnekar, S. S.},
  journal=tcad, 
  title={{OpeNPDN}: A Neural-Network-Based Framework for Power Delivery Network Synthesis}, 
  year={2022},
  volume={41},
  number={10},
  pages={3515--3528},
}

@article{Chhabria22-TODAES,
author = {Chhabria, V. A. and Ahuja, V. and Prabhu, A. and Patil, N. and Jain, P. and Sapatnekar, S. S.},
title = {Encoder-Decoder Networks for Analyzing Thermal and Power Delivery Networks},
year = {2022},
volume = {28},
number = {1},
journal = todaes,
month = dec,
articleno = {3},
numpages = {27},
note = "27 pages"
}

@article{Zhan08,
    author = "Y. Zhan and S. V. Kumar and S. S. Sapatnekar",
    title = "Thermally-Aware Design",
    journal = fnt,
    volume = 2, 
    number = 3,
    pages = "255--370",
    year = 2008,
    month= "March"
}

@article{Su03,
  author={Haihua Su and Sapatnekar, S. S. and Nassif, S. R.},
  journal=tcad,
  title={Optimal decoupling capacitor sizing and placement for standard-cell layout designs}, 
  year={2003},
  volume={22},
  number={4},
  pages={428-436},
}

@article{Popovich08,
  author={Popovich, Mikhail and Sotman, Michael and Kolodny, Avinoam and Friedman, Eby G.},
  journal=tvlsi,
  title={Effective Radii of On-Chip Decoupling Capacitors}, 
  year={2008},
  volume={16},
  number={7},
  pages={894-907},
}

@inproceedings{Karmokar24,
  title="Analyzing the Impact of {FinFET} Self-Heating on the Performance of {RF} Power Amplifiers",
  author={Karmokar, Nibedita and Tam, Sai-Wang and Dinh, Thanh Viet Dinh and Chhabria, Vidya A. and Harjani, Ramesh and Sapatnekar, Sachin S},
  booktitle=iccad,
  year={2024},
  note = "9 pages"
}

@inproceedings{Goplen05,
author = {Goplen, Brent and Sapatnekar, Sachin},
title = {Thermal via placement in {3D ICs}},
year = {2005},
booktitle = ispd,
pages = "167-174",
numpages = {8},
}

@article{Hoang22,
  author={Hoang, Cong Hiep and Azizi, Arad and Fallahtafti, Najmeh and Rangarajan, Srikanth and Radmard, Vahideh and Arvin, Charles and Sikka, Kamal and Schiffres, Scott and Sammakia, Bahgat},
  journal=cpmt,
  title={Design and Thermal Analysis of a {3-D} Printed Impingement Pin Fin Cold Plate for Heterogeneous Integration Application}, 
  year={2022},
  volume={12},
  number={7},
  pages="1091-1099",
}

@article{Lorenzini16,
title = {Embedded single phase microfluidic thermal management for non-uniform heating and hotspots using microgaps with variable pin fin clustering},
journal = ijhmt,
volume = {103},
pages = {1359-1370},
year = {2016},
author = {Daniel Lorenzini and Craig Green and Thomas E. Sarvey and Xuchen Zhang and Yuanchen Hu and Andrei G. Fedorov and Muhannad S. Bakir and Yogendra Joshi},
}

@ARTICLE{Yan25,
  author={Yan, Geyu and Chung, Euichul and Masselink, Erik and Oh, Shane and Zia, Muneeb and Ramakrishnan, Bharath and Oruganti, Vaidehi and Alissa, Husam and Belady, Christian and Im, Yunhyeok and Joshi, Yogendra and Bakir, Muhannad S.},
  journal=cpmt,
  title={Toward {TSV}-Compatible Microfluidic Cooling for {3D ICs}}, 
  year={2025},
  volume={15},
  number={1},
  pages={104-112},
}

@article{Shohel21a,
author = "M. A. {Al Shohel} and V. A. Chhabria and S. S. Sapatnekar",
title = "A New, Computationally Efficient `{B}lech Criterion' for Immortality in General Interconnects",
journal = dac,
year = 2021,
note = "6 pages"
}

@article{Shohel21b,
author = "M. A. {Al Shohel} and V. A. Chhabria and N.~Evmorfopoulos and S.~S.~Sapatnekar",
title = "Analytical Modeling of Transient Electromigration Stress based on Boundary Reflections",
journal = iccad,
year = 2021,
note = "8 pages"
}

@article{Najm20,
author = "F. Najm and V. Sukharev",
title = "Electromigration simulation and design considerations for integrated circuit power grids",
journal = jvstb,
volume = 38, 
number = 6, 
pages = "{063204:1-063204:12}",
month = "Nov/Dec",
year = 2020
}

@article{Najm21,
author = "F. N. Najm",
title = "Equivalent Circuits for Electromigration",
journal = mr,
voume = 123,
pages = {{114200:1--114200:16}},
month = aug,
year = 2021
}

@inproceedings{Shohel23,
  author={M. A. {Al Shohel} and V. A. Chhabria and N. Evmorfopoulos and S. S. Sapatnekar},
  booktitle=iccad,
  title={Frequency-Domain Transient Electromigration Analysis Using Circuit Theory}, 
  year={2023},
  numpages = 8,
  note = "8 pages"
}

@article{Shohel25,
author = "M. A. {Al Shohel} and V. A. Chhabria and N. Evmorfopoulos and S. S. Sapatnekar",
title = "An analytical solution for transient electromigration stress in multisegment straight-line interconnects based on a stress-wave model",
journal = todaes,
volume = 30,
number = 4,
pages = "57:1 -- 57:31",
year = 2025
}

@article{Shen23,
    author = {Shen, Zesheng and Jing, Siyi and Heng, Yiyuan and Yao, Yifan and Tu, K. N. and Liu, Yingxia},
    title = {Electromigration in three-dimensional integrated circuits},
    journal = apr,
    volume = {10},
    number = {2},
    pages = {021309:1-021309:30},
    year = {2023},
    month = {05},
}

@ARTICLE{Xiong14,
  author={Xiong, Hua and Huang, Zhiheng and Conway, Paul},
  journal=tdmr,
  title={Effects of Stress and Electromigration on Microstructural Evolution in Microbumps of Three-Dimensional Integrated Circuits}, 
  year={2014},
  volume={14},
  number={4},
  pages={995-1004},
}

@article{Suhir09,
    author = {Suhir, E.},
    title = {Predictive Analytical Thermal Stress Modeling in Electronics and Photonics},
    journal = amr,
    volume = {62},
    number = {4},
    pages = {040801:1--040801:20},
    year = {2009},
    month = {06},
}

@article{Marella15,
  author={Marella, Sravan K. and Sapatnekar, Sachin S.},
  journal=tvlsi,
  title={A Holistic Analysis of Circuit Performance Variations in {3-D ICs} With Thermal and {TSV}-Induced Stress Considerations}, 
  year={2015},
  volume={23},
  number={7},
  pages={1308-1321},
}

@article{Kteyan21,
title = {Physics-based simulation of stress-induced and electromigration-induced voiding and their interactions in on-chip interconnects},
journal = me,
volume = {247},
pages = {111585:1-111585:7},
year = {2021},
author = {Armen Kteyan and Valeriy Sukharev},
}

@INPROCEEDINGS{Fang12,
  author={Fang, Jianxin and Gupta, Saket and Kumar, Sanjay V. and Marella, Sravan K. and Mishra, Vivek and Zhou, Pingqiang and Sapatnekar, Sachin S.},
  booktitle=iccad,
  title={Circuit reliability: From Physics to Architectures: Embedded tutorial paper}, 
  year={2012},
  pages={243-246},
}

@ARTICLE{Wang25,
  author={Wang, Zhenyu and Nalla, Pragnya Sudershan and Sun, Jingbo and Goksoy, A. Alper and Mandal, Sumit K. and Seo, Jae-sun and Chhabria, Vidya A. and Zhang, Jeff and Chakrabarti, Chaitali and Ogras, Umit Y. and Cao, Yu},
  journal=tcad,
  title={{HISIM}: Analytical Performance Modeling and Design Space Exploration of {2.5D/3D} Integration for {AI} Computing}, 
  year={2025},
  volume={44},
  number={8},
  pages={3208-3221},
}

@INPROCEEDINGS{Krishnan23,
  author={Krishnan, Gokul and Nair, Gopikrishnan Raveendran and Oh, Jonghyun and Anupreetham, Anupreetham and Nalla, Pragnya Sudershan and Hassan, Ahmed and Yeo, Injune and Kasichainula, Kishore and Seo, Jae-sun and Seok, Mingoo and Cao, Yu},
  booktitle=asscc,
  title="{3D-ISC}: A 65nm {3D} Compatible In-Sensor Computing Accelerator with Reconfigurable Tile Architecture for Real-Time {DVS} Data Compression", 
  year={2023},
  note="3 pages"
}

@ARTICLE{Wang25b,
  author={Wang, Weiyang and Ghobadi, Manya},
  journal={IEEE Micro}, 
  title={Spine-Free Networks for {LLM} Training}, 
  year={2025},
}

@article{Mandal22,
  author={Mandal, Sumit K. and Krishnan, Gokul and Goksoy, A. Alper and Nair, Gopikrishnan Ravindran and Cao, Yu and Ogras, Umit Y.},
  journal=jetcas,
  title={{COIN}: Communication-Aware In-Memory Acceleration for Graph Convolutional Networks}, 
  year={2022},
  volume={12},
  number={2},
  pages={472-485},
}

@inproceedings{Mishra13,
author = {Mishra, V. and Sapatnekar, S. S.},
title = {The impact of electromigration in copper interconnects on power grid integrity},
year = {2013},
booktitle = dac,
articleno = {88},
numpages = {6},
note = "6 pages"
}

@book{Trottenberg2000,
  title     = "Multigrid",
  author    = "Trottenberg, U. and Oosterlee, C. W. and Schuller, A.",
  publisher = "Academic Press",
  year      =  2000,
  address   = "San Diego, CA",
}

@article{Zhuo08,
  author={Zhuo, C. and Hu, J. and Zhao, M. and Chen, K.},
  journal={IEEE Transactions on Computer-Aided Design of Integrated Circuits and Systems}, 
  title={Power Grid Analysis and Optimization Using Algebraic Multigrid}, 
  year={2008},
  volume={27},
  number={4},
  pages={738-751},
}

@inproceedings{Feng08,
  author={Feng, Z. and Li, P.},
  booktitle=iccad,
  title={Multigrid on {GPU}: Tackling Power Grid Analysis on parallel {SIMT} platforms}, 
  year={2008},
  pages={647-654},
}

@inproceedings{Su03b, 
author = {Su, H. and Acar, E. and Nassif, S. R.}, 
title = {Power grid reduction based on algebraic multigrid principles}, 
year = {2003},
booktitle = dac,
pages = {109–112},
}

@article{Feng11,
 author={Feng, Zhuo and Zeng, Zhiyu and Li, Peng},
  journal=tvlsi,
  title={Parallel On-Chip Power Distribution Network Analysis on Multi-Core-Multi-{GPU} Platforms}, 
  year={2011},
  volume={19},
  number={10},
  pages={1823-1836},
}

@inproceedings{devgan:iccad00,
        author = "A. Devgan and H. Ji and W. Dai",
        title = "How to Efficiently Capture On-Chip Inductance Effects:
			{I}ntroducing a New Circuit Element {K}",
        booktitle = iccad,
        pages = {150-155},
        year = 2000
}

@article{Ho75,
        author = "C.W. Ho and A.E. Ruehli and P.A. Brennan",
        title = "The Modified Nodal Approach to Network Analysis",
        journal = cas,
        volume = "{CAS-22}",
        number = "6",
        pages = {504--509},
        month = jun,
        year = 1975
}

@inproceedings{Cao02,
        author = "Y. Cao and Y.-M. Lee and T.-H. Chen and C. C.-P. Chen",
	title = "{HiPRIME: H}ierarchical and Passivity Reserved Interconnect Macromodeling Engine for {RLKC} Power Delivery",
	booktitle = dac,
	pages = {379--384},
	year = 2002
}

@INPROCEEDINGS{Su00,
	author = "Haihua Su and Kaushik H. Gala and Sachin S. Sapatnekar",
	title = "Fast Analysis and Optimization of Power/Ground Networks",
	booktitle = iccad,
	pages = {477-480},
	year = 2000
}

@inproceedings{Jiang24,
author = {Jiang, Wenjing and Chhabria, Vidya A. and Sapatnekar, Sachin S.}, 
title = {{IR}-Aware {ECO} Timing Optimization Using Reinforcement Learning}, 
year = {2024},
booktitle = mlcad,
note = "7 pages" 
}

@inproceedings{Kannan20,
  author={Kannan, K. T. and Iyer, Subramanian S.},
  booktitle=ectc,
  title={Deep Trench Capacitors in Silicon Interconnect Fabric}, 
  year={2020},
  pages={2295-2301},
}

@inproceedings{Todri09,
author = "Todri, Aida and Marek-Sadowska, Malgorzata and Maire, Francois and Matheron, Christophe",
booktitle = isqed,
title = {A study of decoupling capacitor effectiveness in power and ground grid networks},
year = {2009},
pages = {653-658},
}

@article{Park20,
  author={Park, Hyunwook and Park, Junyong and Kim, Subin and Cho, Kyungjun and Lho, Daehwan and Jeong, Seungtaek and Park, Shinyoung and Park, Gapyeol and Sim, Boogyo and Kim, Seongguk and Kim, Youngwoo and Kim, Joungho},
  journal=cpmt, 
  title={Deep Reinforcement Learning-Based Optimal Decoupling Capacitor Design Method for Silicon Interposer-Based {2.5-D/3-D ICs}}, 
  year={2020},
  volume={10},
  number={3},
  pages={467-478},
}

@inproceedings{Li22,
author = {Li, Fuping and Wang, Ying and Cheng, Yuanqing and Wang, Yujie and Han, Yinhe and Li, Huawei and Li, Xiaowei}, 
title = {{GIA}: A Reusable General Interposer Architecture for Agile Chiplet Integration}, 
year = {2022}, 
booktitle = iccad,
note = "9 pages"
}

@INPROCEEDINGS{Veloso22,
  author={Veloso, A. and Jourdain, A. and Radisic, D. and Chen, R. and Arutchelvan, G. and O’Sullivan, B. and Arimura, H. and Stucchi, M. and Keersgieter, A. De and Hosseini, M. and Hopf, T. and D’Have, K. and Wang, S. and Dupuy, E. and Mannaert, G. and Vandersmissen, K. and Iacovo, S. and Marien, P. and Choudhury, S. and Schleicher, F. and Sebaai, F. and Oniki, Y. and Zhou, X. and Gupta, A. and Schram, T. and Briggs, B. and Lorant, C. and Rosseel, E. and Hikavyy, A. and Loo, R. and Geypen, J. and Batuk, D. and Martinez, G. T. and Soulie, J. P. and Devriendt, K. and Chan, B. T. and Demuynck, S. and Hiblot, G. and der Plas, G. Van and Ryckaert, J. and Beyer, G. and Litta, E. Dentoni and Beyne, E. and Horiguchi, N.},
  booktitle=svtc, 
  title={Scaled {FinFETs} Connected by Using Both Wafer Sides for Routing via Buried Power Rails}, 
  year={2022},
  pages={284-285},
}

@inproceedings{Hafez23,
  author={Hafez, W. and Agnihotri, P. and Asoro, M. and Aykol, M. and Bains, B. and Bambery, R. and Bapna, M. and Barik, A. and Chatterjee, A. and Chiu, P.C. and Chu, T. and Firby, C. and Fischer, K. and Fradkin, M. and Greve, H. and Gupta, A. and Haralson, E. and Haran, M. and Hicks, J. and Illa, A. and Jang, M. and Klopcic, S. and Kobrinsky, M. and Kuns, B. and Lai, H.-h. and Lanni, G. and Lee, S.-H. and Lindert, N. and Lo, C.-l. and Luo, Y. and Malyavanatham, G. and Marinkovic, B. and Maymon, Y. and Nabors, M. and Neirynck, J. and Packan, P. and Paliwal, A. and Pantisano, L. and Paulson, L. and Penmatsa, P. and Prasad, C. and Puls, C. and Rahman, T. and Ramaswamy, R. and Samant, S. and Sell, B. and Sethi, K. and Shah, F. and Shamanna, M. and Shang, K. and Li, Q. and Sibakoti, M. and Stoeger, J. and Strutt, N. and Thirugnanasambandam, R. and Tsai, C. and Wang, X. and Wang, A. and Wu, S.-j. and Xu, Q. and Zhong, X.-h. and Natarajan, S.},
  booktitle=svtc, 
  title={Intel {PowerVia} Technology: Backside Power Delivery for High Density and High-Performance Computing}, 
  year={2023},
  note="2 pages"
}

@ARTICLE{Lin22,
  author={Lin, H. and van der Plas, G. and Sun, X. and Velenis, D. and Catthoor, F. and Lauwereins, R. and Beyne, E.},
  journal=tvlsi, 
  title={Efficient Backside Power Delivery for High-Performance Computing Systems}, 
  year={2022},
  volume={30},
  number={11},
  pages={1748-1756},
}

@inproceedings{Wong06,
 author = {E. Wong and J. Minz and S. K. Lim},
 title = {Decoupling capacitor planning and sizing for noise and leakage reduction},
 booktitle = iccad,
 year = {2006},
 pages = {395--400},
}

@inproceedings{Gu07Jun,
title={A Switched Decoupling Capacitor Circuit for On-Chip Supply Resonance Damping},
author={J. Gu and H. Eom and C. H. Kim},
booktitle=svc,
year={2007},
pages={126-127}
}

@inproceedings{Xu07,
title={On-Die Supply-Resonance Suppression Using Band-Limited Active Damping},
author={J. Xu and P. Hazucha and M. Huang and P. Aseron and F. Paillet and G. Schrom and J. Tschanz and C. Zhao},
booktitle=isscc,
year={2007},
pages={286-603}
}

@inproceedings{Huang07,
title={Power Delivery for 3{D} Chip Stacks: {P}hysical Modeling and Design Implication},
author={G. Huang and M. Bakir and A. Naeemi and H. Chen and J. D. Meindl},
booktitle=epep,
year={2007},
pages={205-208},
}

@inproceedings{Jain08,
 author = {P. Jain and T. Kim and J. Keane and C. H. Kim},
 title = {A Multi-Story Power Delivery Technique for 3{D} Integrated Circuits},
 booktitle = islped,
 year = {2008},
 pages = {57-62}
}

@inproceedings{Gu05,
 author = {J. Gu and C. H. Kim},
 title = {Multi-Story Power Delivery for Supply Noise Reduction and Low Voltage Operation},
 booktitle = islped,
 year = {2005},
 isbn = {1-59593-137-6},
 pages = {192-197}
 }

@inproceedings{Rajapandian05,
title={High-Tension Power Delivery: {O}perating 0.18 $\mu$m {CMOS} Digital Logic at 5.4{V}},
author={S. Rajapandian and K. Shepard and P. Hazucha and T. Karnik},
booktitle=isscc,
year={2005},
pages={298-599}
}

@inproceedings{Zhou09,
  author = "P. Zhou and S. S. Sapatnekar",
  title = "Congestion-Aware Power Grid Optimization for {3D} Circuits Using
		{MIM} and {CMOS} Decoupling Capacitors",
  booktitle = aspdac,
  year = 2009,
  pages = "179-184"
}

@article{Zhan07,
  author  = "Y. Zhan and S. S. Sapatnekar",
  title   = "High Efficiency {G}reen Function-Based Thermal Simulation
             Algorithms",
  journal = cad,
  year    = "2007",
  month   = "September",
  volume  = "26",
  number  = "9",
  pages   = "1661-1675"
}

@inproceedings{Chhabria20,
author={Chhabria, Vidya A. and Kahng, Andrew B. and Kim, Minsoo and Mallappa, Uday and Sapatnekar, Sachin S. and Xu, Bangqi},
  booktitle=aspdac,
  title={Template-based {PDN} Synthesis in Floorplan and Placement Using Classifier and {CNN} Techniques}, 
  year={2020},
  pages={44-49},
}

@inproceedings{Yang25a,
title = "Adaptive Graph Learning for Efficient Thermal Analysis of Multi-Stacking Chiplet Systems under Interface Variations",
author = "Z. Yang and Y. Cao and J. Sun and V. A. Chhabria",
booktitle = iccad,
year = 2025,
note = "9 pages"
}

@inproceedings{Zhang25,
title = "Optimal Selection and Placement of Voltage Regulators in {2.5D} Heterogeneously Integrated Systems",
author = "H. Zhang and D. Yogi and R. Harjani and S. S. Sapatnekar",
booktitle = iccad,
year = 2025,
note = "9 pages"
}

@inproceedings{Chhabria21,
  author={Chhabria, Vidya A. and Zhang, Yanqing and Ren, Haoxing and Keller, Ben and Khailany, Brucek and Sapatnekar, Sachin S.},
  booktitle=date,
  title={{MAVIREC}: {ML}-Aided Vectored {IR}-Drop Estimation and Classification}, 
  year={2021},
  pages={1825-1828},
}

@inproceedings{Chhabria21BeGAN,
  author={Chhabria, V. A. and Kunal, K. and Zabihi, M. and Sapatnekar, S. S.},
  booktitle=iccad, 
  title={{BeGAN}: Power Grid Benchmark Generation Using a Process-portable {GAN}-based Methodology}, 
  year={2021},
  note="8 pages"
}

@inproceedings{Wu26,
title = "{DALI-PD}: Diffusion-based Synthetic Layout Heatmap Generation for {ML} in Physical Design",
author = "Bing-Yue Wu and Vidya A. Chhabria",
booktitle = aspdac,
year = 2026,
note = "7 pages"
}

@techreport{DoE2023,
author = "Shehabi, Arman and Newkirk, Alex and Smith, Sarah J and Hubbard, Alex and Lei, Nuoa and Siddik, Md Abu Bakar and Holecek, Billie and Koomey, Jonathan and Masanet, Eric and Sartor, Dale",
title = "2024 {United States} Data Center Energy Usage Report",
year = 2024,
institution = "Lawrence Berkeley National Laboratory",
number = "LBNL-2001637",
url = "https://doi.org/10.71468/P1WC7Q"
}

@inproceedings{Krishnakumar24,
  author={Krishnakumar, Sriharini and Choi, Mingeun and Khorasani, Ramin Rahimzadeh and Sharma, Rohit and Swaminathan, Madhavan and Kumar, Satish and Partin-Vaisband, Inna},
  booktitle=ectc,
  title={Vertical Power Delivery for High Performance Computing Systems with Buck-Derived Regulators}, 
  year={2024},
  pages={2136-2142},
}

@article{Avula22,
  author={Avula, Venkatesh and Bhattacharyya, Bidyut and Smet, Vanessa and Joshi, Yogendra and Swaminathan, Madhavan},
  journal=cpmt, 
  title={Multiphysics Challenges and Opportunities for Integrated Voltage Regulators in Power Delivery Architectures}, 
  year={2022},
  volume={12},
  number={1},
  pages={131-146},
}

@ARTICLE{Baek22,
  author={Baek, Jaeil and Elasser, Youssef and Radhakrishnan, Kaladhar and Gan, Houle and Douglas, Jonathan P. and Krishnamurthy, Harish K. and Li, Xin and Jiang, Shuai and Sullivan, Charles R. and Chen, Minjie},
  journal=itpe, 
  title={Vertical Stacked {LEGO-PoL CPU} Voltage Regulator}, 
  year={2022},
  volume={37},
  number={6},
  pages={6305-6322},
}

@book{Rabaey09,
author = "J. Rabaey",
title = "Low Power Design Essentials",
publisher = "Springer",
address = "New York, NY",
year = 2009
}

@INPROCEEDINGS{Abramson21,
  author={Abramson, Rose A. and Ye, Zichao and Ge, Ting and Pilawa-Podgurski, Robert C. N.},
  booktitle=apec,
  title={A High Performance 48-to-6 {V} Multi-Resonant Cascaded Series-Parallel ({CaSP}) Switched-Capacitor Converter},
  year={2021},
  pages={1328-1334},
}

@misc{MPS23,
  title = {{48V Datacenter Solutions: DC/DC Power Conversion for Datacenter, Open Compute, and AI Applications}},
  author = {{Monolithic Power Systems}},
  year = 2023,
  url = "https://media.monolithicpower.com/mps_cms_document/4/8/48v_solution_product_brochure-q3-2023.pdf",
  note = "{Product brochure, Q3 2023. Accessed: July 16th, 2026}"
}

@article{Butzen23,
  author={Butzen, Nicolas and Krishnamurthy, Harish and Ahmed, Zakir and Weng, Sheldon and Ravichandran, Krishnan and Zelikson, Michael and Tschanz, James and Douglas, Jonathan},
  journal={IEEE Solid-State Circuits Letters},
  title={A Monolithic 26 {A/mm$^2$} Continuously Scalable Conversion Ratio Switched-Capacitor Converter with Phase-Merging Turbo and Communication-Less Ganging},
  year={2023},
  volume={6},
  pages={273-276},
}

@article{Lu23,
  author={Lu, Qi and Li, Shuangmu and Zhao, Bo and Jiang, Junmin and Chen, Zhiyuan and Du, Sijun},
  journal=itpe,
  title={A Dynamically Reconfigurable Recursive Switched-Capacitor {DC--DC} Converter with Adaptive Load Ability Enhancement},
  year={2023},
  volume={38},
  number={4},
  pages={5032-5040},
}

@phdthesis{Assem22,
  author={Assem, Pourya},
  title={Integrated Hybrid Switched-Capacitor Converters for Point of Load Power Delivery},
  school={University of California, Berkeley},
  address={Berkeley, CA},
  year={2022},
  note={Tech. Rep. UCB/EECS-2022-243}
}

@article{Krishnakumar26,
  author={Krishnakumar, Sriharini and Popryho, Yaroslav and Choi, Mingeun and Khorasani, Ramin Rahimzadeh and Swaminathan, Madhavan and Kumar, Satish and Partin-Vaisband, Inna},
  journal=cpmt,
  title={A Comprehensive Design Framework for Vertical Power Delivery in High-Performance Computing}, 
  year={2026},
  note="(early access, 14 pages)"
}

@article{Swaminathan26,
  author={Swaminathan, Madhavan and Khorasani, Ramin Rahimzadeh and Partin-Vaisband, Inna and Sharma, Rohit and Kumar, Satish},
  journal=cpmt,
  title={Integrated Vertical Power Delivery – Review \& Challenges}, 
  year={2026},
  note="(early access, 18 pages)"
}

@article{Ye26,
  author={Ye, Zichao and Popovich, Mikhail and Chen, Cheng-Wei and Lu, Jian and Sizikov, Gregory and Gan, Houle},
  journal=cpmt,
  title={Towards {10kAmp AI} Chip: A System Perspective from Lateral to Vertical Power Delivery}, 
  year={2026},
  note="(early access, 12 pages)"
}

@inproceedings{Krishnakumar26ectc,
  author={Krishnakumar, Sriharini and Partin-Vaisband, Inna},
  booktitle=ectc,
  title={Dynamic Power Management Methodology for Distributed Vertical Power Delivery in High-Performance Computing Systems}, 
  year={2026},
  pages={979-984},
}

@article{Choi26,
  author={Choi, Mingeun and Krishnakumar, Sriharini and Popryho, Yaroslav and Khorasani, Ramin Rahimzadeh and Swaminathan, Madhavan and Partin-Vaisband, Inna and Kumar, Satish},
  journal=cpmt,
  title={Self-Consistent Electrothermal Modeling of Distributed Vertical Power Delivery Architecture With Substrate-Embedded Microfluidic Cooling}, 
  year={2026},
  volume={16},
  number={7},
  pages={1534-1542},
}

@article{Choi25,
  author={Choi, Mingeun and Krishnakumar, Sriharini and Rahimzadeh Khorasani, Ramin and Swaminathan, Madhavan and Partin-Vaisband, Inna and Kumar, Satish},
  journal=cpmt,
  title={Substrate-Embedded Microfluidic Cooling of Distributed Vertical Power Delivery Architectures for High-Performance Computing Processors}, 
  year={2025},
  volume={15},
  number={9},
  pages={1912-1920},
}

@misc{Yue26,
  author = "Peiyi Yue and Sachin S. Sapatnekar",
  title = "Intermediate Bus Voltage Selection for Vertical Power Delivery in {2.5D} Integrated Systems",
  note = "(under review)"
}

@IEEEtranBSTCTL{IEEEexample:BSTcontrol,
CTLmax_names_forced_etal = "6"}

@string{aspdac = "Proceedings of the Asia-South Pacific Design Automation	Conference"}

@string{isscc = "Proceedings of the IEEE International Solid-State Circuits	Conference"}

@string{iedm = "IEEE International Electronic Devices Meeting"}

@string{ectc = "Electronics Components and Technology Conference"}

@string{cad = " IEEE T. Comput. Aid D."}

@string{tdmr = " IEEE T. Device Mater. Rel."}

@string{jssc = " IEEE J. Solid-St. Circ."}

@string{iccad = " Proc. ICCAD"}

@string{dac = " Proc. DAC"}

@string{aspdac = " Proc. ASP-DAC"}

@string{cicc = " Proc. CICC"}

@string{mr = " Microelectron. Reliab."}

@string{isqed = " Proc. ISQED"}

@string{ispd = " Proc. ISPD"}

@string{iedm = " Proc. IEDM"}

@string{todaes = "ACM T. Des. Automat. El."}

@string{tcad = "IEEE T. Comput. Aid. D."}

@string{tvlsi = "IEEE T. VLSI Syst"}

@string{cas1 = "IEEE TCAS-I"}

@inproceedings{KWSu03,
author={K. W. Su and Y. M. Sheu and C. K. Lin and S. J. Yang and W. J. Liang and X. Xi and C. S. Chiang and J. K. Her and Y. T. Chia and C. H. Diaz and C. Hu},
title={A scaleable model for {STI} mechanical stress effect on layout dependence of {MOS} electrical characteristics}, 
booktitle=cicc, 
pages={245--248},
year={2003}
}

@ARTICLE{Todri14,
  author={Todri-Sanial, Aida and Kundu, Sandip and Girard, Patrick and Bosio, Alberto and Dilillo, Luigi and Virazel, Arnaud},
  journal=tcad, 
  title={Globally Constrained Locally Optimized 3-D Power Delivery Networks}, 
  year={2014},
  volume={22},
  number={10},
  pages={2131-2144},
}

@string{jssc = "IEEE J. Solid-St. Circ."}

@misc{HIR_2024,
  title = {{Heterogeneous Integration Roadmap}},
  year = 2025,
  url = "https://eps.ieee.org/technology/heterogeneous-integration-roadmap/",
  note ="{Accessed: April 27, 2026}"
}

@INPROCEEDINGS{AMD_3D,
  author={Naffziger, Samuel and Lepak, Kevin and Paraschou, Milam and Subramony, Mahesh},
  booktitle=isscc, 
  title={{AMD} Chiplet Architecture for High-Performance Server and Desktop Products}, 
  year={2020},
  volume={},
  number={},
  pages={44-45},
  doi={10.1109/ISSCC19947.2020.9063103}}

@INPROCEEDINGS{Vertical_PD_Inna,
  author={Krishnakumar, Sriharini and Partin-Vaisband, Inna},
  booktitle=socc, 
  title={Vertical Power Delivery for Emerging Packaging and Integration Platforms -- Power Conversion and Distribution}, 
  year={2023},
  note="6 pages"
}

@inproceedings{LDO_regulator_2014,
  title={Efficient simulation-based optimization of power grid with on-chip voltage regulator},
  author={Yu, Ting and Wong, Martin DF},
  booktitle=aspdac,
  pages={531--536},
  year={2014},
}

@article{VR_Place_2023,
  title={Power Aware Placement of On-Chip Voltage Regulators},
  author={Bairamkulov, Rassul and Friedman, Eby G},
  journal=cad,
  volume={43},
  number={2},
  pages={654--666},
  year={2023},
}

@article{LDOs_2018,
  title={Optimal allocation of {LDOs} and decoupling capacitors within a distributed on-chip power grid},
  author={Sadat, Sayed Abdullah and Canbolat, Mustafa and K{\"o}se, Sel{\c{c}}uk},
  journal=todaes,
  volume={23},
  number={4},
  numpages=15,
  year={2018},
  note="15 pages"
}

@inproceedings{GPU_Acceleration,
  title={Tradeoff analysis and optimization of power delivery networks with on-chip voltage regulation},
  author={Zeng, Zhiyu and Ye, Xiaoji and Feng, Zhuo and Li, Peng},
  booktitle=dac,
  pages={831--836},
  year={2010}
}

@ARTICLE{Zeppelin_3D,
  author={Burd, Thomas and Beck, Noah and White, Sean and Paraschou, Milam and Kalyanasundharam, Nathan and Donley, Gregg and Smith, Alan and Hewitt, Larry and Naffziger, Samuel},
  journal=jssc, 
  title={“{Z}eppelin”: An {SoC} for Multichip Architectures}, 
  year={2019},
  volume={54},
  number={1},
  pages={133-143},
  doi={10.1109/JSSC.2018.2873584}}

@INPROCEEDINGS{Pal21,
  author={Pal, Saptadeep and Liu, Jingyang and Alam, Irina and Cebry, Nicholas and Suhail, Haris and Bu, Shi and Iyer, Subramanian S. and Pamarti, Sudhakar and Kumar, Rakesh and Gupta, Puneet},
  booktitle=dac,
  title={Designing a 2048-Chiplet, 14336-Core Waferscale Processor}, 
  year={2021},
  pages={1183-1188},
}

@INPROCEEDINGS{Gomes22,
  author={Gomes, Wilfred and Koker, Altug and Stover, Pat and Ingerly, Doug and Siers, Scott and Venkataraman, Srikrishnan and Pelto, Chris and Shah, Tejas and Rao, Amreesh and O'Mahony, Frank and Karl, Eric and Cheney, Lance and Rajwani, Iqbal and Jain, Hemant and Cortez, Ryan and Chandrasekhar, Arun and Kanthi, Basavaraj and Koduri, Raja},
  booktitle=isscc,
  title={{Ponte Vecchio}: A Multi-Tile {3D} Stacked Processor for Exascale Computing}, 
  year={2022},
  volume={65},
  pages={42-44},
}

@INPROCEEDINGS{Agarwal22,
  author={Agarwal, Rahul and Cheng, Patrick and Shah, Priyal and Wilkerson, Brett and Swaminathan, Raja and Wuu, John and Mandalapu, Chandrasekhar},
  booktitle=ectc, 
  title={{3D} Packaging for Heterogeneous Integration}, 
  year={2022},
  pages={1103-1107},
}

@INPROCEEDINGS{Prakash24,
  author={Prakash, Pranav Raj and Nabih, Ahmed and Liang, Yan and Kudva, Sudhir and Mosa, Mostafa and Gray, C. Thomas and Li, Qiang},
  booktitle=apec,
  title={A 2400 {W/in$^3$ 1.8 V} Bus Converter Enabling Vertical Power Delivery for Next-Generation Processors}, 
  year={2024},
  pages={910-917},
}

@ARTICLE{Zhou14,
  author={Zhou, Pingqiang and Paul, Ayan and Kim, Chris H. and Sapatnekar, Sachin S.},
  journal=tvlsi,
  title={Distributed On-Chip Switched-Capacitor {DC–DC} Converters Supporting {DVFS} in Multicore Systems}, 
  year={2014},
  volume={22},
  number={9},
  pages={1954-1967},
}

@INPROCEEDINGS{Le13,
  author={Le, Hanh-Phuc and Crossley, John and Sanders, Seth R. and Alon, Elad},
  booktitle=isscc, 
  title={A sub-ns response fully integrated battery-connected switched-capacitor voltage regulator delivering 0.19{W}/mm$^2$ at 73\% efficiency}, 
  year={2013},
  pages={372-373},
}

@INPROCEEDINGS{Harjani14,
  author={Harjani, Ramesh and Chaubey, Saurabh},
  booktitle=cicc, 
  title={A unified framework for capacitive series-parallel {DC-DC} converter design}, 
  year={2014},
  note="8 pages"
}

@inproceedings{Dally23,
author = "Bill Dally",
title = "Hardware for deep learning",
booktitle = hotchips,
year = 2023
}

@inproceedings{Gan24,
  author={Gan, Houle and Jiang, Shuai and Teng, Sue and Yamamoto, Shin and Chivukula, Venkata and Edwards, Bill and Chung, Chee and Chen, Jason and Mohideen, Mushafik and Sizikov, Gregory and Li, Xin},
  booktitle=apec, 
  title={Vertical Power Delivery for 1000 {Amps} Machine Learning {ASICs}}, 
  year={2024},
  pages={906-909},
}

@ARTICLE{Sandri17,
  author={Sandri, Paolo},
  journal={IEEE Power Electron. Mag.}, 
  title={Increasing Hyperscale Data Center Efficiency: A Better Way to Manage {54-V/48-V}-to-Point-of-Load Direct Conversion}, 
  year={2017},
  volume={4},
  number={4},
  pages={58-64},
}

@article{Krein17,
  author={Krein, Philip T.},
  journal={CPSS Trans. Power Electron. Appl.}, 
  title={Data center challenges and their power electronics}, 
  year={2017},
  volume={2},
  number={1},
  pages={39-46},
}

@article{Morra25,
  author = "J. Morra and M. Wood",
  title = "Disaggregating power in data centers",
  year = 2025,
  journal = "Electron. Des.",
  month = may,
  day = 12
}

@article{Ahmed21,
author={Ahmed, Mohamed H. and Lee, Fred C. and Li, Qiang},
  journal={IEEE J. Emerging Sel. Top. Power Electron.}, 
  title={Two-Stage {48-V VRM} With Intermediate Bus Voltage Optimization for Data Centers}, 
  year={2021},
  volume={9},
  number={1},
  pages={702-715},
}

@article{Kong25,
  author={Kong, Cai and Huan, Weiwei and Wang, Jian and Li, Dali and Sun, Hui and Zhang, Xuehong and Ye, Fenghua and Lin, Kaizhi},
  journal=itpe, 
  title={A Vertical Power Delivery Architecture for High-Performance Computing}, 
  year={2025},
  volume={40},
  number={5},
  pages={6663-6674},
}

@inproceedings{Sapatnekar09,
  author={Sapatnekar, Sachin S.},
  booktitle=aspdac, 
  title={Addressing thermal and power delivery bottlenecks in {3D} circuits}, 
  year={2009},
  pages={423-428},
}

@INPROCEEDINGS{Su:dac02,
	author = "Haihua Su and Jiang Hu and Sani R. Nassif and Sachin S. Sapatnekar",
	title = "Congestion-driven codesign of power and signal networks",
	booktitle = dac,
	pages = {477-480},
	address = "New Orleans, LA",
	month = jun,
	year = 2002
}

@article{Zhao02,
        author = "Min Zhao and Rajendran V. Panda and Sachin S. Sapatnekar and
			David Blaauw",
        title = "Hierarchical Analysis of Power Distribution Networks",
        journal = cad,
        volume = 21,
        number = 2,
        pages = {159--168},
	month = feb,
        year = 2002 
}

@article{Kozhaya02,
        author = "Joseph Kozhaya and Sani R. Nassif and Farid N. Najm",
        title = "A Multigrid-like Technique for Power Grid Analysis",
        journal = cad,
        volume = 21,
        number = 10,
        pages = {1148--1160},
	    month = oct,
        year = 2002 
}

@ARTICLE{Wang17,
  author={Wang, Longfei and Khatamifard, S. Karen and Uzun, Orhun Aras and Karpuzcu, Ulya R. and Köse, Selçuk},
  journal=tvlsi, 
  title={Efficiency, Stability, and Reliability Implications of Unbalanced Current Sharing Among Distributed On-Chip Voltage Regulators}, 
  year={2017},
  volume={25},
  number={11},
  pages={3019-3032},
}

@INPROCEEDINGS{Gupta07,
  author={Gupta, Meeta S. and Oatley, Jarod L. and Joseph, Russ and Wei, Gu-Yeon and Brooks, David M.},
  booktitle=date, 
  title={Understanding Voltage Variations in Chip Multiprocessors using a Distributed Power-Delivery Network}, 
  year={2007},
  note = "6 pages"
}

@inproceedings{Tang25,
title = "Thermal Challenges and Opportunities in Components to System Integration for {AI}",
author = "Weihua Tang and Madhusudan Iyengar",
booktitle = "IEEE HI Workshop at ECTC",
year = 2025
}

@article{Tan03,
  author={Tan, S. X. D. and Shi, C.-J. R. and Jyh-Chwen Lee},
  journal=tcad, 
  title={Reliability-constrained area optimization of VLSI power/ground networks via sequence of linear programmings}, 
  year={2003},
  volume={22},
  number={12},
  pages={1678-1684},
}

@ARTICLE{Singh05,
  author={Singh, J. and Sapatnekar, S.S.},
  journal=tcad, 
  title={Congestion-aware topology optimization of structured power/ground networks}, 
  year={2005},
  volume={24},
  number={5},
  pages={683-695},
}

\end{document}